\documentclass[12pt,onecolumn]{IEEEtran}

\usepackage{amsmath,amssymb,epsfig,psfrag,cite,subfigure}
\include{macros}
\usepackage{graphicx}
\usepackage{epstopdf}

\usepackage{nicefrac}

\usepackage{lipsum,multicol}

\usepackage{algorithm}
\usepackage{algpseudocode}

\usepackage{enumitem}
\usepackage{mwe}
\usepackage{epsfig,psfrag}
\usepackage{subfigure}
\usepackage{color}
\usepackage{url}
\usepackage{mathtools,xparse}

\usepackage{gensymb}

\allowdisplaybreaks

\newtheorem{proof}{Proof}

\newcommand{\mtA}{{\mathcal{A}}}

\newcommand{\mtE}{{\mathcal{E}}}
\newcommand{\mtP}{{\mathcal{P}}}
\newcommand{\mtPtilde}{{ \widetilde{\mathcal{P}} }}
\newcommand{\mtPs}{{ \mathcal{P}_{s} }}

\newcommand{\expectation}{{\mathbb{E}}}

\newcommand{\Jboldi}{{ \boldsymbol{\mathrm{J}}^{-1} }}
\newcommand{\Jbold}{{ \boldsymbol{\mathrm{J}} }}

\newcommand{\Gammabold}{{ \boldsymbol{\mathrm{\Gamma}} }}
\newcommand{\gammabold}{{ \boldsymbol{\mathrm{\gamma}} }}

\newcommand{\Gammaboldhat}{{ \boldsymbol{\mathrm{\widehat{\Gamma}}} }}
\newcommand{\Gammabolddelta}{{ \boldsymbol{\mathrm{\Delta} \mathrm{\Gamma}} }}

\newcommand{\Abold}{{ \boldsymbol{\mathrm{A}} }}
\newcommand{\Bbold}{{ \boldsymbol{\mathrm{B}} }}
\newcommand{\Cbold}{{ \boldsymbol{\mathrm{C}} }}

\newcommand{\Phibold}{{ \boldsymbol{\mathrm{\Phi}} }}

\newcommand{\Hbold}{{ \boldsymbol{\mathrm{H}} }}

\newcommand{\Imatrix}{{ \boldsymbol{\mathrm{I}} }}

\newcommand{\Pt}{{ P_{\mathrm{T}} }}

\newcommand{\Et}{{ \mtI_{\mathrm{ind}} }}
\newcommand{\Etilde}{{ \widetilde{\mtI} }}

\newcommand{\boldzero}{{ {\boldsymbol{0}} }}
\newcommand{\boldone}{{ {\boldsymbol{1}} }}

\newcommand{\boldonetr}{{ \boldone^\transpose }}

\newcommand{\phibold}{{ {\boldsymbol{\mathrm{\phi}}} }}
\newcommand{\varphibold}{{ {\boldsymbol{\mathrm{\varphi}}} }}

\newcommand{\avgE}{{ \mtI_{\rm{avg}} }}
\newcommand{\avgEtilde}{{ \widetilde{\mtI}_{\rm{avg}} }}

\newcommand{\Pavgstar}{{ P^{\star}_{\rm{avg}} }}

\newcommand{\pp}{{ {\mathbf{p}} }}
\newcommand{\pplb}{{ \mathbf{p}_{\mathrm{lb}} }}
\newcommand{\ppub}{{ \mathbf{p}_{\mathrm{ub}} }}
\newcommand{\ppstar}{{ {\mathbf{p}^{\star}} }}

\newcommand{\ppast}{{ {\mathbf{p}^{*}} }}

\newcommand{\xx}{{ {\mathbf{x}} }}

\newcommand{\boldonet}{{ \boldone^\transpose }}
\newcommand{\trace}{{ {\mathrm{trace}} }}

\newcommand{\gammaik}{{ \gamma^{(i)}_{k_1,k_2} }}
\newcommand{\gammaikgen}[1]{{ \gamma^{(#1)}_{k_1,k_2} }}
\newcommand{\gammaiksyn}{{ \gamma^{(i),\rm{syn}}_{k_1,k_2} }}
\newcommand{\gammaikasyn}{{ \gamma^{(i),\rm{asy}}_{k_1,k_2} }}

\newcommand{\transpose}{{ T }}
\newcommand{\realset}[1]{{ \mathbb{R}^{#1} }}

\newcommand{\Xbold}{{ \boldsymbol{\mathrm{X}} }}
\newcommand{\Xboldi}{{ \boldsymbol{\mathrm{X}}^{-1} }}

\newcommand{\NL}{{N_{\rm{L}}}}

\newcommand{\lr}{{\boldsymbol{l}_{\mathrm{r}}}}
\newcommand{\lrh}{{\hat{\boldsymbol{l}}_{\mathrm{r}}}}

\newcommand{\lrer}{{ {\mathbf{e}}_{\lr} }}

\newcommand{\lt}[1]{{\boldsymbol{l}^{#1}_{\mathrm{t}}}}
\newcommand{\RR}{\mathbb{R}}

\newcommand{\nr}{{\boldsymbol{n}_{\mathrm{r}}}}
\newcommand{\nt}[1]{{\boldsymbol{n}^{#1}_{\mathrm{t}}}}

\newcommand{\lrs}[1]{{l_{\mathrm{r},#1}}}
\newcommand{\lts}[2]{{l^{#1}_{\mathrm{t},#2}}}

\newcommand{\nrs}[1]{{n_{\mathrm{r},#1}}}
\newcommand{\nts}[2]{{n^{#1}_{\mathrm{t},#2}}}

\DeclarePairedDelimiter{\norm}{\lVert}{\rVert}

\newcommand{\lrhat}{{\hat{\boldsymbol{l}}_{\mathrm{r}}}}

\newcommand{\lrerset}{{ \mathcal{E}_{\lr} }}
\newcommand{\ee}{{ {\mathbf{e}} }}

\newcommand{\deltalr}{{ \delta_{\lr} }}
\newcommand{\deltatheta}{{ \delta_{\theta} }}
\newcommand{\deltaphi}{{ \delta_{\phi} }}

\newcommand{\thetahat}{{ \hat{\theta} }}
\newcommand{\phihat}{{ \hat{\phi} }}

\newcommand{\etheta}{{ e_{\theta} }}
\newcommand{\ephi}{{ e_{\phi} }}

\newcommand{\thetaset}{{ \mathcal{E}_{\theta} }}
\newcommand{\phiset}{{ \mathcal{E}_{\phi} }}

\newcommand{\egeneralset}{{ \mathcal{\widetilde{E}} }}
\newcommand{\egeneral}{{ {\mathbf{\widetilde{e}}} }}

\newcommand{\funcgeneral}{{ \psi(\pp,\egeneral) }}
\newcommand{\funcgeneralstar}{{ \psi(\ppstar,\egeneral) }}
\newcommand{\funcgeneralstarr}{{ \psi(\ppstar,\egeneral_{\miter+1}) }}

\newcommand{\preg}{{ \varrho }}
\newcommand{\miter}{{ n }}
\newcommand{\Ngrid}{{ N_{\rm{grid}} }}

\newcommand{\bigO}{{ \mathcal{O} }}

\newcommand{\funcgeneralout}{{ \Psi(\pp) }}
\newcommand{\funcgeneraloutm}{{ \Psi^{\miter}(\pp) }}
\newcommand{\funcgeneraloutmp}{{ \Psi_{\preg}^{\miter}(\pp) }}
\newcommand{\funcgeneraloutmpstar}{{ \Psi_{\preg}^{\miter}(\ppstar) }}
\newcommand{\egeneralsetm}{{ \mathcal{\widetilde{E}}_{\miter} }}

\newcommand{\tildest}{{ \widetilde{s}_i(t) }}
\newcommand{\Tsi}{{ T_{s,i} }}

\newcommand{\Pelec}{{ E_i^{\rm{elec}} }}
\newcommand{\Popt}{{ E_i^{\rm{opt}} }}

\newcommand{\Popttilde}{{ \widetilde{E}_i^{\rm{opt}} }}

\newcommand{\mtI}{{ \mathcal{I} }}
\newcommand{\Etind}{{ \mtI^{i}_{\mathrm{ind}} }}

\newcommand{\fci}{{ f_{c,i} }}

\newcommand{\ppnonrobust}{{ \pp^{\rm{n-rob}} }}

\begin{document}

\title{Optimal and Robust Power Allocation for Visible Light Positioning Systems under Illumination Constraints}

\author{Musa Furkan Keskin,\thanks{M. F. Keskin, A. D. Sezer, and S. Gezici are with the Department of Electrical and Electronics Engineering, Bilkent University, 06800, Ankara, Turkey, Tel: +90-312-290-3139, Fax: +90-312-266-4192, Emails: \{keskin,adsezer,gezici\}@ee.bilkent.edu.tr.} Ahmet Dundar Sezer, and Sinan Gezici\vspace{-0.3cm}}

\maketitle

\begin{abstract}
	The problem of optimal power allocation among light emitting diode (LED) transmitters in a visible light positioning (VLP) system is considered for the purpose of improving localization performance of visible light communication (VLC) receivers. Specifically, the aim is to minimize the Cram\'{e}r-Rao lower bound (CRLB) on the localization error of a VLC receiver by optimizing LED transmission powers in the presence of practical constraints such as individual and total power limitations and illuminance constraints. The formulated optimization problem is shown to be convex and thus can efficiently be solved via standard tools. We also investigate the case of imperfect knowledge of localization parameters and develop robust power allocation algorithms by taking into account both overall system uncertainty and individual parameter uncertainties related to the location and orientation of the VLC receiver. In addition, we address the total power minimization problem under predefined accuracy requirements to obtain the most energy-efficient power allocation vector for a given CRLB level. Numerical results illustrate the improvements in localization performance achieved by employing the proposed optimal and robust power allocation strategies over the conventional uniform and non-robust approaches.

\end{abstract}

\section{Introduction}
\subsection{Background and Motivation}
With the advent of low-cost and energy-efficient light emitting diode (LED) technologies, LED based visible light communication (VLC) systems have gathered a significant amount of research interest in the last decade \cite{BeyondPoint,SurveyVLC15,Jovicic}. Utilizing the vast unlicensed visible light spectrum, VLC has the potential to surmount the issue of spectrum scarcity encountered in radio frequency (RF) based wireless systems \cite{VLC_Survey}. In indoor scenarios, VLC systems can employ the available lighting infrastructure to provide various capabilities simultaneously, such as illumination, high-speed data transmission, and localization \cite{SurveyVLC15,VLP_Roadmap}. Apart from their basic function of illuminating indoor spaces, LEDs can be modulated at high frequencies to accomplish high data rate transmission \cite{Rapid_VLC,Jovicic,HighSpeedVLC}. On the other hand, the process of localization via visible light signals can be realized by visible light positioning (VLP) systems, where VLC receivers equipped with photo detectors can perform position estimation by exploiting signals emitted by LED transmitters at known locations \cite{HighSpeedVLC,VLP_Roadmap,panta2012indoor,IndoorVisLig}. Since line-of-sight (LOS) links generally exist between LED transmitters and VLC receivers, and multipath effects are not very significant as compared to RF based solutions \cite{ghassemlooy2017visible,Fundamental_VLC_2004}, VLP systems can facilitate precise location estimation in indoor environments \cite{IndoorVisLig,EPSILON,zhang2014asynchronous,LED_MultiRec}.


In order to provide satisfactory performance for mobile or stationary devices, it is essential to investigate performance optimization in visible light systems with respect to various criteria, such as mean-squared error (MSE) minimization (e.g., \cite{MIMO_VLC_JSAC_2015,MIMO_VLC_Design_TCOM_2017,Dimming_MIMO_VLC_2016,JTOD_TWC_2017}) and transmission rate maximization (e.g., \cite{Power_Offset_VLC_TCOM_2013,ResourceAlloc_JLT_2014,PowRateOpt_VLC_TSP_2015, DCO_OFDM_TSP_2016,Guvenc_JSAC_2017,RateOpt_VLC_JLT_2016,NOMA_VLC_2017,PowAlloc_VLC_JSAC_2017,VLC_Precoding_TCOM_2017}). In the literature, transmit precoding and DC offset\footnote{Optical intensity modulation in VLC systems requires that the amplitude of the electrical drive current of the LED must be non-negative \cite{Power_Offset_VLC_TCOM_2013}.} designs are extensively explored to improve the MSE performance of multiple-input multiple-output (MIMO) VLC systems \cite{MIMO_VLC_Design_TCOM_2017,JTOD_TWC_2017,MIMO_VLC_JSAC_2015,Dimming_MIMO_VLC_2016}. In addition to transceiver and offset designs in VLC systems, an increasingly popular research strand focuses on power allocation for LED transmitters to enhance system performance \cite{ResourceAlloc_JLT_2014,RateOpt_VLC_JLT_2016,PowRateOpt_VLC_TSP_2015,DCO_OFDM_TSP_2016,Power_Offset_VLC_TCOM_2013,NOMA_VLC_2017,Guvenc_JSAC_2017,PowAlloc_VLC_JSAC_2017,PowAlloc_VLC_2012,PowAlloc_JLT_2017}. Due to practical concerns related to energy efficiency and LED lifespan, transmission powers of LEDs in visible light systems are valuable resources that can have profound effects on both transmission rates of VLC systems and localization accuracy of VLP systems. In \cite{Power_Offset_VLC_TCOM_2013}, the total instantaneous data rate of LED arrays is considered as the performance metric for a MIMO VLC system and the optimal strategy for LED power allocation is derived under sum optical power and non-negativity constraints. The studies in \cite{ResourceAlloc_JLT_2014} and \cite{DCO_OFDM_TSP_2016} perform power optimization for LEDs to maximize the sum transmission rate of all subcarriers in a VLC system employing optical orthogonal frequency-division multiplexing (OFDM). With the aim of achieving proportional fairness among users in a multi-user VLC network, the total logarithmic throughput is optimized in \cite{Guvenc_JSAC_2017} and \cite{PowAlloc_VLC_JSAC_2017} to identify the optimal LED power control strategy. Although total and individual power constraints are extensively utilized in power allocation optimization in VLC systems, several studies incorporate color and luminance constraints into the power optimization framework, as well, in compliance with the illumination functionality of VLC systems \cite{PowRateOpt_VLC_TSP_2015,RateOpt_VLC_JLT_2016}. In general, power allocation algorithms in both VLC and VLP systems should take into account a variety of design requirements imposed by the multi-faceted nature of visible light applications.

The concept of power allocation has also been widely considered for RF based wireless localization networks \cite{Robust_Pow_Alloc_Win_2013,Win_2014_IEEEnetw_PowerOpt,Win_PA_Coop_2015,Joint_Alloc_TCOM_2016,PA_Game_TSP_2016,PA_OFDM_TWC_2017}, where the transmit powers of anchor nodes (the locations of which are known) can be optimized to improve the localization accuracy of target nodes (with unknown locations). The prevailing approach in such investigations is to adopt a mathematically tractable and tight bound on the localization error as the performance metric and to formulate the optimization problem under average and peak anchor power constraints. In \cite{Robust_Pow_Alloc_Win_2013} and \cite{Win_2014_IEEEnetw_PowerOpt}, anchor power allocation algorithms are designed to minimize the total power consumption subject to predefined accuracy requirements for localization of target nodes. For cooperative localization networks, distributed power allocation strategies are developed in \cite{Win_PA_Coop_2015}, where the transmit powers of both anchors and targets are optimally allocated to minimize the squared position error bound (SPEB). Moreover, \cite{PA_OFDM_TWC_2017} explores the problem of optimal power allocation for OFDM subcarriers in the presence of both perfect and imperfect knowledge of network parameters. As commonly observed in RF wireless localization systems, optimal power allocation provides non-negligible performance benefits over the traditional uniform strategy for a wide range of localization scenarios.

\subsection{Contributions}
Motivated by the promising performance improvements achieved via power allocation in both RF localization networks and VLC systems, we propose the problem of optimal power allocation for LED transmitters in a VLP system, where the objective is to minimize the localization error of the VLC receiver subject to practical constraints related to power and illumination. Leveraging tools from convex optimization and semidefinite programming (SDP), we formulate and solve various optimization problems in both the absence and presence of parameter uncertainties. The power allocation problem for VLP systems has the following key differences from the one in RF based localization systems: \textit{(i)} Due to the limited linear region of operation, the LEDs are subject to both the minimum and peak power constraints \cite{LED_Linearity,PowRateOpt_VLC_TSP_2015,VLC_Lighting_Mag,Dimming_MIMO_VLC_2016}. \textit{(ii)} Since VLP systems serve the dual purpose of illumination and localization, the problem formulation should include lighting constraints that guarantee an acceptable level of illumination in indoor spaces \cite{VLC_ComMag_2014,VLC_Guvenc_Lighting,Lighting_JLT_2008,VLC_Lighting_Mag}. \textit{(iii)} In contrast to RF systems in which multipath components can severely affect the quality of localization, the received signal power in VLP systems can accurately be characterized by the Lambertian formula \cite{Fundamental_VLC_2004}.

The main contributions of this study can be listed as follows:
\begin{itemize}[leftmargin=*]
	\item \textit{Problem Formulation for LED Power Allocation:} For the first time in the literature, we investigate the problem of optimal power allocation among LED transmitters in a VLP system for maximizing the localization accuracy of a VLC receiver. Specifically, we employ the Cram\'{e}r-Rao lower bound (CRLB) on the localization error as the performance measure and formulate the power allocation problem to minimize the CRLB in the presence of transmission power and illumination constraints.
	\item \textit{Robustness Under Overall System Uncertainty\footnote{Overall system uncertainty is defined as the uncertainty related to all the system parameters except for the transmit powers and mathematically formulated as a perturbation matrix.}:} We consider the problem of robust power allocation under imperfect knowledge of system parameters and demonstrate that the resulting worst-case CRLB minimization problem can equivalently be transformed into a convex program, which further simplifies to an SDP via constraint relaxation.
	\item \textit{Robustness Under Individual Parameter Uncertainties:} We present robust power allocation schemes in the presence of uncertainties in the location and orientation of the VLC receiver. To tackle the resulting intractable optimization problems, we propose an iterative entropic regularization approach where, at each iteration, a convex problem is solved and a three (two)-dimensional grid search is executed over the uncertainty region corresponding to the location (orientation) of the VLC receiver.
	\item \textit{Sum Power Minimization Under Preset Accuracy Constraints:} We formulate the minimum power consumption problem under the constraint that the CRLB for localization of the VLC receiver does not exceed a certain threshold. We also extend the problem to the case of overall system uncertainty and prove that the resulting worst-case accuracy constrained optimization problem is shown to be reformulated as a convex one, leading to efficient solutions.
\end{itemize}
In addition, numerical results show that the proposed optimal power allocation approach for LED transmitters yields significant localization performance gains over the conventional uniform power assignment method. For the case of imperfect knowledge of localization related parameters, the proposed robust power allocation strategies are shown to outperform the uniform and non-robust (which disregards the uncertainty in parameter measurements) strategies.



\section{System Model}\label{sec:SysModel}

We consider a VLP system with $\NL$ LED transmitters and a VLC receiver, where the objective is to estimate the unknown location of the VLC receiver by utilizing signals emitted by the LED transmitters. As is commonly the case for visible light systems \cite{VLP_Roadmap,CRB_TOA_VLC}, we assume a line-of-sight (LOS) scenario between each LED transmitter and the VLC receiver. Then, the received signal at the VLC receiver due to the $i$th LED transmitter is formulated as \cite{CRB_TOA_VLC}
\begin{gather}\label{eq:recSig1}
r_i(t)=\alpha_i R_p \, s_i(t-\tau_i) + \eta_i(t)
\end{gather}
for $i \in \{1,\ldots,\NL\}$ and $t\in[T_{1,i},T_{2,i}]$, where $T_{1,i}$ and $T_{2,i}$ specify the observation interval for the signal coming from the $i$th LED transmitter,
$\alpha_i$ is the optical channel attenuation factor between the $i$th LED transmitter and the VLC receiver ($\alpha_i>0$), $R_p$ is the responsivity of the photo detector at the VLC receiver, $s_i(t)$ is the transmitted signal of the $i$th LED transmitter, which is nonzero over an interval of $[0,T_{s,i}]$, $\tau_i$ is the TOA of the signal emitted by the $i$th LED transmitter, and $\eta_i(t)$ is zero-mean additive white Gaussian noise with a spectral density level of $\sigma^2$. To enable independent processing of signals coming from different LED transmitters, a certain type of multiple access protocol, such as frequency-division or time-division multiple access \cite{book_goldsmith,multiaccessVLP}, can be employed \cite{VLC_Survey}. Thus, the noise processes, $\eta_1(t),\ldots,\eta_{\NL}(t)$, are modeled to be independent. In addition, we assume that the VLC receiver has the knowledge of $R_p$ and $s_i(t)$, $i \in \{1,\ldots,\NL\}$.

The TOA parameter in \eqref{eq:recSig1} can be expressed as
\begin{gather}\label{eq:tau}
\tau_i = \norm{\lr - \lt{i}}\big/c+\Delta_i 
\end{gather}
where $c$ is the speed of light, $\Delta_i$ denotes the time offset between the clocks of the $i$th LED transmitter and the VLC receiver,
$\lr = \left[\lrs{1}\,\, \lrs{2}\,\, \lrs{3} \right]^T$ and $\lt{i} = \left[\lts{i}{1} \,\, \lts{i}{2} \,\, \lts{i}{3} \right]^T$ denote the locations of the VLC receiver and the $i$th LED transmitter, respectively, and $\norm{\lr - \lt{i}}$ specifies the distance between the $i$th LED transmitter and the VLC receiver. For synchronous VLP systems, $\Delta_i = 0$ for $i=1,\ldots,\NL$, whereas for asynchronous systems, $\Delta_i$'s are modeled as deterministic unknown parameters. 

Using the Lambertian model \cite{WirelessInfComm_97}, the channel attenuation factor $\alpha_i$ in \eqref{eq:recSig1} can be written as
\begin{gather}\label{eq:alpha}
\alpha_i 
= -\frac{(m_i+1) S}{2\pi} \frac{\left[(\lr - \lt{i})^{T} \nt{i} \right]^{m_i} (\lr - \lt{i})^{T} \nr}{\norm{\lr - \lt{i}}^{m_i+3}}   
\end{gather}
where $m_i$ is the Lambertian order for the $i$th LED transmitter, $S$ is the area of the photo detector at the VLC receiver, and $\nr = \left[\nrs{1}\,\, \nrs{2}\,\, \nrs{3} \right]^T$ and $\nt{i} = \left[\nts{i}{1} \,\, \nts{i}{2} \,\, \nts{i}{3} \right]^T$ specify the orientation vectors of the VLC receiver and the $i$th LED transmitter, respectively \cite{CRB_TOA_VLC,Guvenc_hybrid}.\footnote{For example, $\nr = \left[0\,\, 0\,\, 1 \right]^T$ means that the VLC receiver is pointing upwards.}

It is assumed that the parameters $S$, $\nr$, $m_i$, $\lt{i}$, and $\nt{i}$ for $i=1,\ldots,\NL$ are known by the VLC receiver. For example, the orientation of the VLC receiver, $\nr$, can be measured through a gyroscope and the parameters of the LED transmitters ($m_i$, $\lt{i}$ and $\nt{i}$) can be transmitted to the receiver via visible light communications.

\section{Optimal Power Allocation for LEDs}\label{sec:optLED}
In this section, we establish a theoretical framework for the optimization of LED transmit powers with the aim of maximizing the localization performance of the VLC receiver. First, we describe the optimization variables and the optimization performance metric. Then, by incorporating several practical constraints related to indoor visible light scenarios, we present the formulation of the optimal power allocation problem.

\subsection{Optimization Variables}\label{sec:opt_var}
The transmitted signal $s_i(t)$ from the $i$th LED transmitter can be expressed as
\begin{equation}\label{eq:opt_var_sit}
s_i(t) = \sqrt{P_i}\, \tildest
\end{equation}
for $i \in \{1,\ldots,\NL\}$, where $\tildest$ is a base signal such that $\int_{0}^{\Tsi} (\tildest)^2 dt/\Tsi = 1$, and $P_i$ is a parameter that specifies the transmit power of the $i$th LED. Then, the \textit{optical power} of $s_i(t)$ can be obtained as \cite{CRB_TOA_VLC}
\begin{align} \label{eq:p_opt}
\Popt = \int_{0}^{\Tsi} s_i(t) dt \big/\Tsi = \sqrt{P_i} \, \Popttilde
\end{align}
where $\Tsi$ denotes the period of $s_i(t)$ and $\Popttilde$ is the optical power of $\tildest$, defined as
\begin{align}\label{eq:p_opt_tilde}
\Popttilde &\triangleq \int_{0}^{\Tsi} \tildest dt\big/\Tsi~.
\end{align}
On the other hand, the \textit{electrical power} consumed by the $i$th LED, $\Pelec$, is proportional to $P_i$ \cite{WirelessInfComm_97}; that is, $\Pelec \propto \int_{0}^{\Tsi} (s_i(t))^2 dt/\Tsi = P_i$.
In this study, we aim at optimizing the electrical powers of the transmitted signals by adjusting $\{P_i\}_{i=1}^{\NL}$ to maximize the localization performance.

\subsection{Optimization Metric}\label{sec:metric}
The CRLB on the variance of any unbiased estimator $\lrh$ for the location of the VLC receiver $\lr$ can be expressed as
\vspace{-0.2in}
\begin{gather}\label{eq:CRLBgen}
\expectation \big\{ \|\lrh - \lr \|^2 \big\}\geq\trace\big\{\Jboldi(\pp) \big\}
\end{gather}
where the Fisher information matrix (FIM) is given by \cite{Direct_TCOM}
\begin{align}\label{eq:FIM_compact}
\Jbold(\pp) = (\Imatrix_3 \otimes \pp)^\transpose \Gammabold
\end{align}\vspace{-0.3in}
with
\begin{align}\label{eq:Hmatrix}
\pp &\triangleq \left[ P_1 \, \ldots \, P_{\NL} \right]^\transpose \in \realset{\NL} \\ \label{eq:Hmatrix2}
\Gammabold &\triangleq \begin{bmatrix}
\gammabold_{1,1} & \gammabold_{1,2}  & \gammabold_{1,3}  \\
\gammabold_{2,1} & \gammabold_{2,2} & \gammabold_{2,3} \\
\gammabold_{3,1} & \gammabold_{3,2} & \gammabold_{3,3}
\end{bmatrix} \in \realset{3 \NL \times 3} \\ \label{eq:gamma_vec}
\gammabold_{k_1,k_2} &\triangleq \left[ \gammaikgen{1} \,\ldots\, \gammaikgen{\NL} \right]^\transpose \in \realset{\NL}
\end{align}
for $k_1,k_2\in\{1,2,3\}$. $P_i$ in \eqref{eq:Hmatrix} is as defined in Section~\ref{sec:opt_var}, $\gammaikgen{i}$ in \eqref{eq:gamma_vec} is given by Appendix~\ref{app_FIM} and $\otimes$ in \eqref{eq:FIM_compact} represents the Kronecker product.

We employ the CRLB in \eqref{eq:CRLBgen} as the optimization performance metric for quantifying the localization accuracy of the VLC receiver. The reason for this choice is that the maximum likelihood (ML) estimator for the location of the VLC receiver can attain the CRLB for sufficiently high signal-to-noise ratios (SNRs) \cite{Poor,VanTrees}. In addition, the CRLB expression facilitates theoretical analyses and results in mathematically tractable formulations. Also, the usage of the CRLB as a performance measure renders the analysis independent of any specific location estimator, thereby providing a generic framework for power allocation in VLP systems.

\subsection{VLP System Constraints}\label{sec:constraints}
Certain constraints must be imposed on a VLP system while designing LED power optimization schemes in order to satisfy illumination, energy, and hardware related requirements. In particular, the following system constraints are taken into account in the power optimization problem:

\subsubsection{Individual Power Constraints}
Lower and upper bound constraints for LED powers must be incorporated to ensure that transmission powers of LEDs lie inside the linear region of operation so that the LED output power is proportional to the input drive current, which provides efficient electrical-to-optical conversion \cite{LED_Linearity,PowRateOpt_VLC_TSP_2015,OFDM_VLC_2014,VLC_Lighting_Mag,Dimming_MIMO_VLC_2016}. In addition, self-heating induced by a high drive current may shorten the LED lifetime \cite{Lampe_VLC_TCOM_2015}. Hence, the resulting constraint set is given by
\begin{align}\label{eq:cons_x1}
\mtP_1 \triangleq \{ \pp \in \realset{\NL}: \pplb \preceq \pp \preceq \ppub  \}
\end{align}
where $\pplb \in \realset{\NL}$ and $\ppub \in \realset{\NL}$ denote, respectively, the lower and upper bounds on $\pp$ in \eqref{eq:Hmatrix}.
	
\subsubsection{Total Power Constraint}
Due to power consumption restrictions of LEDs and human eye safety considerations, the total electrical power of LEDs in a VLP system must be limited \cite{WirelessInfComm_97,Clipping_OFDM_TCOM_2012,DCO_OFDM_TSP_2016,VLC_Lighting_Mag}. Therefore, we have the following constraint set regarding the total power limit:
\begin{align}\label{eq:cons_x2}
\mtP_2 \triangleq \{ \pp \in \realset{\NL}:  \boldone^\transpose \pp \leq \Pt \}
\end{align}
where $\Pt$ determines the total electrical power constraint of LEDs.

\subsubsection{Individual Illumination Constraints}
Since VLP systems are utilized also for indoor lighting in addition to other benefits such as data transmission and localization, a certain level of brightness must be maintained over the room and/or at specified locations \cite{VLC_ComMag_2014,VLC_Guvenc_Lighting,Lighting_JLT_2008,VLC_Lighting_Mag}. The illuminance ($\rm{lm/m^2},\rm{lx}$) is used as a measure of brightness, which is defined as the luminous flux ($\rm{lm}$) per unit area \cite{LED_Book}. Combining \cite[Eq.~3]{Lighting_JLT_2008}, \cite[Eq.~16.3]{LED_Book} and \eqref{eq:p_opt}, the horizontal illuminance at location $\xx$ generated by the $i$th LED can be calculated as
\vspace{-0.45cm}\begin{align}
\Etind(\xx,P_i) = \sqrt{P_i} \, \phi_{i}(\xx)
\end{align}
where
\vspace{-0.4cm}\begin{align}\label{eq:phi}
\phi_{i}(\xx) \triangleq \frac{(m_i+1) \kappa_i \Popttilde}{2\pi} \frac{\left[(\xx - \lt{i})^\transpose \nt{i} \right]^{m_i} (\lts{i}{3} - x_{3}) }{\norm{\xx - \lt{i}}^{m_i+3}}
\end{align}
with $\Popttilde$ being as defined in \eqref{eq:p_opt_tilde} and $\kappa_i$ denoting the luminous efficacy ($\rm{lm/W}$) of the $i$th LED, defined as the optical power to luminous flux conversion efficiency \cite{LED_Book}. Then, the total illuminance at $\xx$ produced by all the LEDs can be obtained as follows \cite{Illuminance_Sum}:
\begin{align}\label{eq:ill_total}
\Et(\xx,\pp) = \sum_{i=1}^{\NL} \Etind(\xx,P_i) = \sum_{i=1}^{\NL} \sqrt{P_i} \, \phi_{i}(\xx)
\end{align}
Let $L$ denote the number of locations at which the illuminance constraint is to be satisfied. Then, the corresponding constraint set can be defined as
\begin{align}\label{eq:cons_x3}
\mtP_3 \triangleq \{ \pp \in \realset{\NL}:  \Et(\xx_{\ell},\pp) \geq \Etilde_{\ell}, ~ \ell = 1,\ldots,L\}
\end{align}
where $\Etilde_{\ell}$ is the illuminance constraint defined for location $\xx_{\ell}$.

\subsubsection{Average Illumination Constraint}
The expression in \eqref{eq:ill_total} quantifies the illuminance level at a specified location in the room. It may also be necessary to keep the average illuminance over the room above a certain threshold to comply with average brightness requirements. Then, the average illuminance is 
\begin{align}\label{eq:ill_avg}
\avgE(\pp) = \sum_{i=1}^{\NL} \sqrt{P_i} \, \frac{\int_{\mtA} \phi_i(\xx) d \xx}{|\mtA|}
\end{align}
where $\mtA$ denotes the region where the average illuminance constraint must be satisfied and $|\mtA|$ denotes the volume of $\mtA$. The constraint set associated with the average illuminance is given by
\begin{align}\label{eq:cons_x4}
\mtP_4 \triangleq \{ \pp \in \realset{\NL}: \avgE(\pp) \geq \avgEtilde \}
\end{align}
where $\avgEtilde$ is the average illuminance constraint.
	
\vspace{-0.4cm}

\subsection{Problem Formulation}
Considering the optimization metric in Section~\ref{sec:metric} and the system constraints in Section~\ref{sec:constraints}, the problem of optimal power allocation for LED transmitters can be formulated as follows:
\begin{subequations}\label{eq:optLED}
\begin{align} \label{eq:optLED_obj}
\mathop{\mathrm{minimize}}\limits_{\pp} &~~
\trace \big\{ \Jboldi(\pp) \big\}  \\ \label{eq:optLED_ind_power}
\mathrm{subject~to}&~~ \pp \in \mtP
\end{align}
\end{subequations}
where $\mtP \triangleq \bigcap_{i=1}^{4} \mtP_i$ and $\Jbold(\pp) $ is given by \eqref{eq:FIM_compact}. In the proposed power optimization framework in \eqref{eq:optLED}, we search for the optimal power vector that minimizes the CRLB for the localization of the VLC receiver subject to power and illumination constraints. The following lemma establishes the convexity of \eqref{eq:optLED}.

\textbf{Lemma~1.} \textit{The optimization problem in \eqref{eq:optLED} is convex.}

\begin{proof}
First, the convexity of $f(\pp) \triangleq \trace \big\{ \Jboldi(\pp) \big\}$ in $\pp$ is shown as follows: Consider any $\pp_1 \in \realset{\NL}$, $\pp_2 \in \realset{\NL}$, and $\lambda \in [0, 1]$. Then,
	\begin{align}\label{eq:trace1}
	f(\lambda \pp_1 + (1-\lambda) \pp_2) &= \mathrm{trace} \Big\{ \Big( \big[\Imatrix_3 \otimes (\lambda \pp_1 + (1-\lambda) \pp_2)\big]^\transpose \Gammabold \Big)^{-1} \Big\} \\ \label{eq:trace2}
	&= \trace \Big\{ \Big(\lambda (\Imatrix_3 \otimes \pp_1)^\transpose \Gammabold + (1-\lambda) (\Imatrix_3 \otimes \pp_2)^\transpose \Gammabold \Big)^{-1} \Big\} \\ \label{eq:trace3}
	& \leq \lambda f(\pp_1) + (1-\lambda) f(\pp_2)
	\end{align}
where \eqref{eq:trace1} follows from \eqref{eq:FIM_compact}, \eqref{eq:trace2} is the result of the properties of Kronecker product, and \eqref{eq:trace3} is due to the convexity of $\trace \big\{ \Xboldi \big\}$ for $\Xbold \succ 0$ \cite{boyd_convex}. Since the constraint sets $\mtP_1$ in \eqref{eq:cons_x1} and $\mtP_2$ in \eqref{eq:cons_x2} are linear, and $\mtP_3$ in \eqref{eq:cons_x3} and $\mtP_4$ in \eqref{eq:cons_x4} are convex due to the concavity of \eqref{eq:ill_total} and \eqref{eq:ill_avg} with respect to $\pp$, the combined constraint set $\mtP$ is convex, thus proving the convexity of \eqref{eq:optLED} in $\pp$.
\end{proof}

Based on Lemma~1, it is noted that optimal LED power allocation strategies can be obtained via standard convex optimization tools \cite{cvx2014,boyd_convex}.

\section{Robust Power Allocation with Overall System Uncertainty}\label{sec:robust_overall}
In Section~\ref{sec:optLED}, the optimal power allocation is performed by assuming perfect knowledge of localization parameters, which however may not be realistic for practical VLP scenarios. In this section, robust optimization schemes will be designed for power allocation among LED transmitters in the presence of \textit{overall uncertainty} in VLP system parameters\footnote{The meaning of overall uncertainty will be clarified in Section~\ref{sec:overall_uncertainty}.}. In the following, we present the problem formulation for robust power allocation with overall system uncertainty in VLP scenarios and demonstrate that it can be reformulated as a convex optimization problem, which can further be simplified to an SDP via feasible set relaxations.

\subsection{Problem Statement}\label{sec:overall_uncertainty}
Considering the optimization problem in \eqref{eq:optLED}, the matrix $\Gammabold$ in \eqref{eq:Hmatrix2} contains all the information required for LED power optimization based on \eqref{eq:FIM_compact}. Since the knowledge of localization related parameters is imperfect in practice, it is assumed that $\Gammabold$ is measured with some uncertainty; that is,
\begin{equation}\label{eq:matrix_error_model}
\Gammaboldhat = \Gammabold + \Gammabolddelta
\end{equation}
where $\Gammaboldhat$ is the estimated/nominal matrix and $\Gammabolddelta$ represents the error matrix that accumulates all the uncertainties in localization parameters. As in \cite{Eldar_Robust_TSP_2005,Robust_Linear_2008,Palomar_TSP_MIMO_2009,Robust_DRSS_TSP_2017}, a deterministically bounded error model is considered for $\Gammabolddelta$, i.e.,
\begin{equation}\label{eq:Gamma_delta_norm}
\Gammabolddelta \in \mtE \triangleq \{ \Gammabolddelta \in \mathbb{R}^{3\NL \times 3}: \norm{\Gammabolddelta} \leq \delta \}
\end{equation}
for a known size of uncertainty region $\delta$, where $\norm{\cdot}$ stands for the matrix spectral norm.

For the robust counterpart of \eqref{eq:optLED}, the aim is to minimize the worst-case CRLB over all uncertainties in the form of $\norm{\Gammabolddelta} \leq \delta$. Hence, considering the error model in \eqref{eq:matrix_error_model}, the robust min-max design problem corresponding to the CRLB optimization in \eqref{eq:optLED} can be stated as follows:
	\begin{equation}\label{eq:optLED_robust_obj}
	\mathop{\mathrm{minimize}}\limits_{\pp} ~ \mathop{\mathrm{max}}\limits_{\Gammabolddelta \in \mtE} ~	\trace \Big\{ \big((\Imatrix_3 \otimes \pp)^\transpose (\Gammaboldhat - \Gammabolddelta)\big)^{-1} \Big\}  ~~~
	\mathrm{subject~to}~~ \pp \in \mtP
	\end{equation}
where $\mtE$ is as defined in \eqref{eq:Gamma_delta_norm} and $\mtP$ is the feasible region in \eqref{eq:optLED_ind_power}.

\subsection{Equivalent Convex Reformulation of \eqref{eq:optLED_robust_obj}}
The problem in \eqref{eq:optLED_robust_obj} is challenging to solve in its current form and its direct solution is computationally prohibitive. In the following proposition, we demonstrate that \eqref{eq:optLED_robust_obj} can be reformulated as a convex program.

\textbf{Proposition~1.} \textit{The robust power allocation problem in \eqref{eq:optLED_robust_obj} can equivalently be expressed as the following convex optimization problem:}
\begin{subequations}\label{eq:optLED_robust_sdp}
\begin{align} \label{eq:optLED_robust_sdp_obj}
\mathop{\mathrm{minimize}}\limits_{\pp,t,\Hbold,s,\mu} &~~ t
\\ \label{eq:optLED_robust_sdp_trace_ineq}
\mathrm{subject~to}&~~ \trace \big\{ \Hbold \big\} \leq t - d s
\\ \label{eq:optLED_robust_sdp_lmi}
&~~ \Phibold(\pp,\Hbold,s,\mu) \succeq 0
\\ \label{eq:optLED_robust_sdp_nonneg_H}
&~~ \Hbold \succeq 0,~\mu \geq 0
\\ \label{eq:optLED_robust_P}
&~~ \pp \in \mtP
\end{align}
\end{subequations}
\textit{where}
\vspace{-0.2cm}\begin{align}\label{eq:phi_sdp}
\Phibold(\pp,\Hbold,s,\mu) \triangleq \begin{bmatrix}
\Hbold + s \Imatrix  & \Imatrix & \boldzero \\
\Imatrix & (\Imatrix_3 \otimes \pp)^\transpose \Gammaboldhat - \mu \Imatrix & - \frac{\delta}{2} (\Imatrix_3 \otimes \pp)^\transpose \\
\boldzero & - \frac{\delta}{2} (\Imatrix_3 \otimes \pp) & \mu \Imatrix
\end{bmatrix}
\end{align}
\textit{and $d$ is the dimension of localization.}

\begin{proof}
	We utilize the following lemmas for the proof \cite{Robust_Linear_2008}.
	
	\textbf{Lemma~2} \textit{(18c in \cite{LecturesConvBook}).} \textit{Let $\Xbold \in \realset{d \times d}$ be a symmetric matrix. Then, $\trace \big\{ \Xbold \big\} \leq t$
if and only if there exists $s \in \realset{}$ and $\Hbold \in  \realset{d \times d}$ such that}
		\begin{align}\label{eq:lemma4_ineq}
		\trace \big\{ \Hbold \big\} \leq t - d s,~
		\Hbold \succeq 0,~
		\Hbold + s \Imatrix \succeq \Xbold ~.
		\end{align}

	
	\textbf{Lemma~3} \textit{(Lemma~2 in \cite{Eldar_Robust_TSP_2005}).} \textit{For matrices $\Abold$, $\Bbold$ and $\Cbold$ with $\Abold = \Abold^\transpose$, the matrix inequality}
	\begin{align}\label{eq:matrix_ineq_lemma5}
	\Abold \succeq \Bbold^\transpose \Xbold \Cbold + \Cbold^\transpose \Xbold^\transpose \Bbold \,, ~~ \forall \Xbold : \norm{\Xbold} \leq \delta
	\end{align}
	\textit{is satisfied if and only if there exists a $\mu \geq 0$ such that}
	\begin{align}
	\begin{bmatrix}
	\Abold - \mu \Cbold^\transpose \Cbold  & - \delta \Bbold^\transpose \\
	- \delta \Bbold & \mu \Imatrix
	\end{bmatrix} \succeq 0 ~.
	\end{align}
	
	By introducing a slack variable $t$, \eqref{eq:optLED_robust_obj} can equivalently be written in the epigraph form as follows:
	\begin{subequations}\label{eq:optLED_robust_eq}
	\begin{align} \label{eq:optLED_robust_eq_obj}
	\mathop{\mathrm{minimize}}\limits_{\pp,t} &~~ t
	\\ \label{eq:trace_inf}
	\mathrm{subject~to}&~~
	\trace \Big\{ \big((\Imatrix_3 \otimes \pp)^\transpose (\Gammaboldhat - \Gammabolddelta)\big)^{-1} \Big\} \leq t \,, \forall \Gammabolddelta: \Gammabolddelta \in \mtE
	\\ &~~ \pp \in \mtP
	\end{align}
	\end{subequations}
	First, using Lemma~2 for the constraint in \eqref{eq:trace_inf} leads to the following inequalities:
	\begin{subequations}
	\begin{align}
	\trace \big\{ \Hbold \big\} &\leq t - d s,~	\Hbold \succeq 0 \\ \label{eq:lemma4_use}
	\Hbold + s \Imatrix &\succeq \big((\Imatrix_3 \otimes \pp)^\transpose (\Gammaboldhat - \Gammabolddelta)\big)^{-1} \,,~ \forall \Gammabolddelta: \Gammabolddelta \in \mtE
	\end{align}
	\end{subequations}
	for some $s \in \realset{}$ and $\Hbold \in  \realset{d \times d}$. Next, applying the Schur complement lemma to \eqref{eq:lemma4_use}, we have
	\begin{align}\label{eq:schur_1}
	\begin{bmatrix}
	\Hbold + s \Imatrix  & \Imatrix \\
	\Imatrix & (\Imatrix_3 \otimes \pp)^\transpose (\Gammaboldhat - \Gammabolddelta)
	\end{bmatrix} \succeq 0 \,,~ \forall \Gammabolddelta: \Gammabolddelta \in \mtE~.
	\end{align}
	Rearranging \eqref{eq:schur_1}, an inequality of the form \eqref{eq:matrix_ineq_lemma5} is obtained as
	\begin{align}\label{eq:schur_2}
	\begin{bmatrix}
	\Hbold + s \Imatrix  & \Imatrix \\
	\Imatrix & (\Imatrix_3 \otimes \pp)^\transpose \Gammaboldhat
	\end{bmatrix} \succeq \Bbold^\transpose \Gammabolddelta \, \Cbold + \Cbold^\transpose \Gammabolddelta^\transpose \Bbold	\,,~   \forall \Gammabolddelta: \Gammabolddelta \in \mtE
	\end{align}
	where $\Bbold \triangleq \frac{1}{2} \left[ \boldzero ~~ (\Imatrix_3 \otimes \pp) \right]$ and $\Cbold \triangleq \left[ \boldzero ~~ \Imatrix \right]$. Then, via Lemma~3, \eqref{eq:schur_2} is transformed into the constraint in \eqref{eq:optLED_robust_sdp_lmi}, which completes the proof.
\end{proof}

\subsection{SDP Formulation via Feasible Set Relaxation}\label{sec:sdp_relax}
Since \eqref{eq:optLED_robust_sdp_lmi} is a linear matrix inequality (LMI) in the variables $\pp$, $\Hbold$, $s$ and $\mu$ \cite{boyd1994linear}, the problem in \eqref{eq:optLED_robust_sdp} is convex. In addition, if the general convex constraint \eqref{eq:optLED_robust_P} can be relaxed to a linear one by replacing $\mtP$ with an appropriate $\mtPtilde$ satisfying $\mtPtilde \supseteq \mtP$, \eqref{eq:optLED_robust_sdp} simplifies to an SDP with a linear objective and a set of LMI constraints \cite{SDP_Boyd_1996}. By squaring both sides of \eqref{eq:cons_x3} and applying the arithmetic mean-geometric mean inequality, a relaxed version of $\mtP_3$ is obtained as
\begin{equation}
\mtPtilde_3 \triangleq \{ \pp : \phibold(\xx)^\transpose \pp \geq \Etilde_{\ell}^2/\boldonetr \phibold(\xx) , ~ \ell = 1,\ldots,L \} \supseteq \mtP_3
\end{equation}
where $\phibold(\xx) \triangleq \left[ \phi_1(\xx) \ldots \phi_{\NL}(\xx) \right]$. Similarly, $\mtP_4$ in \eqref{eq:cons_x4} can be relaxed to
\begin{equation}
\mtPtilde_4 \triangleq \{ \pp : \varphibold^\transpose \pp \geq \avgEtilde^2/\boldonetr \varphibold \} \supseteq \mtP_4
\end{equation}
where $\varphibold \in \realset{\NL}$ with $\varphi_i \triangleq \frac{\int_{\mtA} \phi_i(\xx) d \xx}{|\mtA|} $. Hence, by defining $\mtPtilde \triangleq \mtP_1 \cap \mtP_2 \cap \mtPtilde_3 \cap \mtPtilde_4$ and replacing $\mtP$ with $\mtPtilde$ in \eqref{eq:optLED_robust_P}, \eqref{eq:optLED_robust_sdp} becomes an SDP problem and thus can be solved very efficiently using available convex optimization softwares \cite{cvx2014,yalmip}. The worst-case complexity of an SDP with $n$ variables and $m$ constraints is given by $\bigO(\max(m,n)^4 n^{1/2} \log(1/\epsilon) )$, where $\epsilon$ is the tolerance level \cite{SDP_2010_zhi}. Thus, the computational complexity of the SDP version of \eqref{eq:optLED_robust_sdp}, which is obtained from the feasible set relaxations, can be expressed as $\bigO(N^{4.5}_{\rm{L}} \log(1/\epsilon) )$.


\section{Robust Power Allocation with Individual Parameter Uncertainties}\label{sec:robust_ind}

In this section, we consider robust power allocation schemes under individual uncertainties related to localization parameters in VLP systems. In indoor tracking applications, VLC receiver position $\lr$ can be predicted to lie in a validation region, but its exact position cannot perfectly be known. Similarly, VLC receiver orientation $\nr$ may be subject to measurement errors since the measurement devices such as gyroscopes tend to generate noisy parameter estimates. Hence, individual parameter uncertainties must be taken into account while deriving optimal strategies for LED power allocation. In the following, we first present the problem formulations in the presence of uncertainties in the location and the orientation of the VLC receiver. Then, we propose an iterative approach to solve the resulting intractable optimization problems.

\subsection{Uncertainty in VLC Receiver Location}
To formulate the robust power allocation problem in the presence of uncertainties about the location of the VLC receiver, we assume that the nominal location $\lrh$ is a perturbed version of the true location $\lr$, i.e.,
\begin{equation}\label{eq:lrhat}
\lrh = \lr + \lrer \,.
\end{equation}
As in \cite{Robust_Linear_2008,SourceLoc_TSP_2011,Lampe_MISO_TSP_2016}, we assume a spherical uncertainty set for the location errors, i.e.,
\begin{equation}\label{eq:error_loc}
\lrer \in \lrerset \triangleq \{ \ee \in \RR^{3} : \norm{\ee} \leq \deltalr \}
\end{equation}
where $\deltalr$ is a known value that represents the size of the uncertainty region. Then, the power allocation problem in \eqref{eq:optLED} based on worst-case CRLB minimization can be formulated as
\begin{align} \label{eq:optLED_robust_loc_obj}
\mathop{\mathrm{minimize}}\limits_{\pp} ~~ \mathop{\mathrm{max}}\limits_{ \lrer \in \lrerset } &~~
\trace \Big\{ \big((\Imatrix_3 \otimes \pp)^\transpose \, \Gammabold(\lrh - \lrer) \big)^{-1} \Big\}  \\ \nonumber
\mathrm{subject~to}&~~ \pp \in \mtP
\end{align}
where $\Gammabold(\lrh - \lrer)$ denotes the matrix $\Gammabold$ in \eqref{eq:Hmatrix2} evaluated at $\lr = \lrh - \lrer$.

\subsection{Uncertainty in VLC Receiver Orientation}
The orientation vector of the VLC receiver can be expressed as
\begin{align}\label{eq:nr_angle}
\nr(\theta,\phi) = \left[ \sin \theta \cos \phi ~ \sin \theta \sin \phi ~ \cos \theta \right]^\transpose
\end{align}
where $\theta$ and $\phi$ represent the polar and the azimuth angles, respectively \cite{Lampe_MISO_TSP_2016}. According to \eqref{eq:nr_angle}, the uncertainty related to the orientation of the VLC receiver can be modeled as angular uncertainties in $\theta$ and $\phi$ \cite{Lampe_MISO_TSP_2016}. Hence, the nominal (measured) polar and azimuth angles can be written as
\begin{align}\label{eq:angle_hat}
\thetahat = \theta + \etheta,~~
\phihat = \phi + \ephi
\end{align}
where $\theta$ and $\phi$ are the true values of the angles, and $\etheta$ and $\ephi$ represent the errors in angular measurements, for which the bounded uncertainty sets can be defined as
\begin{subequations}\label{eq:error_angle}
\begin{align} \label{eq:error_angle_theta}
e_{\theta} \in \thetaset &\triangleq \{ e \in \mathbb{R} : |e| \leq \delta_{\theta} \} \\ \label{eq:error_angle_phi}
e_{\phi} \in \phiset &\triangleq \{ e \in \mathbb{R} : |e| \leq \delta_{\phi} \}
\end{align}
\end{subequations}
with $\deltatheta$ and $\deltaphi$ denoting the maximum possible angular deviations. Then, the robust counterpart of \eqref{eq:optLED} in the case of orientation uncertainties can be stated as
\begin{align} \label{eq:optLED_robust_orient_obj}
\mathop{\mathrm{minimize}}\limits_{\pp} \, \mathop{\mathrm{max}}\limits_{\substack{ \etheta \in \thetaset \\ \ephi \in \phiset } } &~
\trace \Big\{ \Big((\Imatrix_3 \otimes \pp)^\transpose \, \Gammabold\big(\nr(\thetahat - \etheta, \phihat - \ephi)\big) \Big)^{-1} \Big\}  \\ \nonumber
\mathrm{subject~to}&~ \pp \in \mtP
\end{align}
where $\nr(\cdot\,,\cdot)$ is as defined in \eqref{eq:nr_angle} and $\Gammabold\big(\nr(\theta,\phi)\big)$ is the matrix $\Gammabold$ in \eqref{eq:Hmatrix2} evaluated at $\nr = \nr(\theta,\phi)$.

\subsection{Iterative Entropic Regularization Algorithm}
In this part, we develop a unified power allocation algorithm design for solving the robust optimization problems in \eqref{eq:optLED_robust_loc_obj} and \eqref{eq:optLED_robust_orient_obj}. To this end, let the error vectors and the corresponding uncertainty sets in \eqref{eq:error_loc} and \eqref{eq:error_angle} be defined as follows:
\begin{align}\label{eq:error_unif}
\egeneral &\triangleq
\begin{cases}
\lrer \,, & \text{uncertainty in VLC receiver location} \\
(\etheta,\ephi) \,, & \text{uncertainty in VLC receiver orientation}
\end{cases}
\\
\label{eq:set_unif}
\egeneralset &\triangleq
\begin{cases}
\lrerset \,, & \text{uncertainty in VLC receiver location} \\
\thetaset \times \phiset \,, & \text{uncertainty in VLC receiver orientation}
\end{cases}
\end{align}
In addition, the objective functions in \eqref{eq:optLED_robust_loc_obj} and \eqref{eq:optLED_robust_orient_obj} can be represented by
\begin{equation}\label{eq:func_unif}
\funcgeneral \triangleq
\begin{cases}
\trace \Big\{ \big((\Imatrix_3 \otimes \pp)^\transpose \, \Gammabold(\lrh - \lrer) \big)^{-1} \Big\}  &  \\
\trace \Big\{ \Big((\Imatrix_3 \otimes \pp)^\transpose \, \Gammabold\big(\nr(\thetahat - \etheta, \phihat - \ephi)\big) \Big)^{-1} \Big\}  &
\end{cases}
\end{equation}
where the first and second rows denote, respectively, the cases for the uncertainty in the location and the orientation. Then, based on \eqref{eq:error_unif}--\eqref{eq:func_unif}, the problems in \eqref{eq:optLED_robust_loc_obj} and \eqref{eq:optLED_robust_orient_obj} can be unified into a single optimization framework as follows:
\begin{align} \label{eq:optLED_robust_unif_obj}
\mathop{\mathrm{minimize}}\limits_{\pp} ~ \mathop{\mathrm{max}}\limits_{ \egeneral \in \egeneralset } ~
\funcgeneral  ~~~
\mathrm{subject~to}&~ \pp \in \mtP
\end{align}

The inner problem in \eqref{eq:optLED_robust_unif_obj} is not convex since $\funcgeneral$ is not concave in $\egeneral$. Moreover, the epigraph form of \eqref{eq:optLED_robust_unif_obj} results in a semi-infinite optimization problem whose constraints (in the form of $\funcgeneral \leq t,\, \forall \, \egeneral \in \egeneralset$, for some $t \in \RR$) do not admit a tractable convex reformulation, as in \eqref{eq:trace_inf}. Furthermore, the exhaustive search method for solving \eqref{eq:optLED_robust_unif_obj} has a computational complexity that is exponential in the number of LED transmitters $\NL$. Therefore, it is challenging to solve \eqref{eq:optLED_robust_unif_obj} in a computationally efficient manner via conventional techniques.

To tackle the robust design problem in \eqref{eq:optLED_robust_unif_obj}, our algorithmic approach is to use an \textit{iterative entropic regularization} procedure that successively decreases the objective value of the outer problem by updating the power vector $\pp$ while simultaneously refining the optimal value of the inner maximization problem \cite{IterEntrReg_2004,SmartGrid_iterativeEntReg_2014}. Let the objective function of the outer problem in \eqref{eq:optLED_robust_unif_obj} be defined as
\begin{equation}\label{eq:func_general}
\funcgeneralout \triangleq \mathop{\mathrm{max}}\limits_{ \egeneral \in \egeneralset } ~ \funcgeneral 	~.
\end{equation}
The continuous uncertainty set $\egeneralset$ can be discretized using $\miter$ points in $\egeneralset$ to obtain a subset $\egeneralsetm$ of $\egeneralset$. Then, $\funcgeneralout$ in \eqref{eq:func_general} can be approximated by $\funcgeneraloutm \triangleq \mathop{\mathrm{max}}\limits_{ \egeneral \in \egeneralsetm } ~ \funcgeneral$. To circumvent the non-differentiability of $\funcgeneraloutm$, we can employ the following entropic regularized/smoothed approximation of the max function \cite{IterEntrReg_2004}, \cite[p.~72]{boyd_convex}:
\begin{equation}\label{eq:entropy_func}
\funcgeneraloutmp \triangleq \frac{1}{\preg} \log \Bigg\{ \sum_{\egeneral \in \egeneralsetm} \exp \big( \preg \, \funcgeneral \big) \Bigg\}
\end{equation}
where $\preg$ is the regularization constant \cite{SmartGrid_iterativeEntReg_2014}.

Based on the regularized function in \eqref{eq:entropy_func}, we propose the iterative entropic regularization algorithm in Algorithm~\ref{alg:ier}, which consists of the following steps \cite{IterEntrReg_2004, SmartGrid_iterativeEntReg_2014}:
\begin{itemize}
	\item \textit{Outer Minimization:} The objective function $\funcgeneralout$ in \eqref{eq:func_general} is approximated by the smoothed version $\funcgeneraloutmp$ in \eqref{eq:entropy_func}. The resulting convex problem\footnote{Since $\funcgeneral$ is a convex function of $\pp$ for a given $\egeneral$ (see \eqref{eq:func_unif} and Lemma~1) and the log-sum-exp function is convex \cite[p.~72]{boyd_convex}, the resulting composition $\funcgeneraloutmp$ is convex in $\pp$.} in \eqref{eq:opt_entropic} can efficiently be solved via standard tools of convex optimization \cite{boyd_convex}.
	\item \textit{Inner Maximization:} Using the power vector $\ppstar$ obtained from the outer minimization step, the inner maximization problem of \eqref{eq:optLED_robust_unif_obj} is solved in \eqref{eq:inner_max} by performing a three (two)-dimensional grid search over $\egeneralset$ for the case of the uncertainty in the location (orientation) of the VLC receiver.
\end{itemize}
Algorithm~\ref{alg:ier} can be shown to converge to a global minimum of \eqref{eq:optLED_robust_unif_obj} \cite{IterEntrReg_2004}. It is worth noting that the computational burden of \eqref{eq:optLED_robust_unif_obj} is significantly reduced via Algorithm~\ref{alg:ier} as compared to the exhaustive search approach, as mentioned in the next subsection.

\begin{algorithm}\footnotesize
	\caption{Iterative Entropic Regularization Algorithm to Solve the Robust Power Allocation Problem in \eqref{eq:optLED_robust_unif_obj}}
	\label{alg:ier}

	\begin{algorithmic}
		\State \textbf{Initialization:} \\ Select $\egeneral_1 \in \egeneralset$, set $\egeneralset_1 = \{ \egeneral_1 \}$, $\miter=1$ and $k=1$. \\ Select $\preg > 0$, $\epsilon \in (0,1)$, $\varsigma > 0$ and $\Ngrid \in \mathbb{Z}^{+}$.

		\State \textbf{Iterative Step:}
		\State \textit{(Outer Problem)} Solve the following convex optimization problem with a tolerance level of $\epsilon^k$:
		\begin{equation}\label{eq:opt_entropic}
		\ppstar = \arg \, \min_{\pp \in \mtP} ~ \funcgeneraloutmp
		\end{equation}
		where $\funcgeneraloutmp$ is given by \eqref{eq:entropy_func}.
		
		\State \textit{(Inner Problem)} Obtain a new candidate from the uncertainty region $\egeneralset$ using a grid search over the prespecified $\Ngrid$ points:
		\begin{equation}\label{eq:inner_max}
		\egeneral_{\miter+1} = \arg \, \max_{\egeneral \in \egeneralset} ~ \funcgeneralstar
		\end{equation}
		where $\funcgeneral$ is as defined in \eqref{eq:func_unif}.
		
		\State Update $k = k+1$.
		\State \textit{(Check the Objective Value)}
		\If{$\funcgeneralstarr > \funcgeneraloutmpstar$}
		\State Set $\egeneralset_{n+1} = \egeneralset_{n} \cup \{ \egeneral_{n+1} \}$.
		\State Update $\miter = \miter + 1$.
		\State Update $\preg = \max(\preg, \log(\miter)^2)$.
		\EndIf
		\State \textit{(Check the Tolerance Value)}
		\If{$\epsilon^k + \log(\miter)/\preg > \varsigma$}
		\State Update $\preg = \preg + \log(\miter)$.
		\EndIf
		
		\State \textbf{Stopping Criteria:} \\
		$\funcgeneralstarr \leq \funcgeneraloutmpstar$ and $\epsilon^k + \log(\miter)/\preg \leq \varsigma$.
		
	\end{algorithmic}
	\normalsize
\end{algorithm}


\subsection{Complexity Analysis}
In this part, we discuss the computational complexity of Algorithm~\ref{alg:ier} and compare it with that of the exhaustive search based solution of \eqref{eq:optLED_robust_unif_obj}. At each iteration, Algorithm~\ref{alg:ier} involves solving a convex optimization problem and a grid search over the uncertainty region. Let $\bigO(C)$ denote the complexity of the convex optimization problem in \eqref{eq:opt_entropic} and $\Ngrid$ the number of points employed for the grid search over $\egeneralset$ in \eqref{eq:inner_max}. Then, the per-iteration complexity of Algorithm~1 is given by $\bigO(C) + \bigO(\Ngrid)$. Regarding the exhaustive search technique for solving \eqref{eq:optLED_robust_unif_obj}, let each axis of the feasible region $\mtP \subset \RR^{\NL}$ be discretized using $\bigO(M)$ different values. Thus, the outer iteration of \eqref{eq:optLED_robust_unif_obj} has a computational complexity of $\bigO(M^{\NL})$. Utilizing $\Ngrid$ points for the inner iteration, the overall complexity becomes $\bigO(M^{\NL} \, \Ngrid)$. Therefore, the complexity of the exhaustive search method grows exponentially with the number of LED transmitters, whereas that of Algorithm~\ref{alg:ier} is primarily determined by the convex problem in \eqref{eq:opt_entropic}, which can be solved in polynomial time \cite{LecturesConvBook}. As a result, Algorithm~\ref{alg:ier} has significantly lower computational complexity than the exhaustive search based solution.

\vspace{-0.3cm}
\section{Minimum Power Consumption Problem}\label{sec:minPowCons}
In practical indoor VLP systems, the power consumption of LEDs and the localization error of VLC receivers must be jointly considered in a power optimization problem. In Section~\ref{sec:optLED}, Section~\ref{sec:robust_overall} and Section~\ref{sec:robust_ind}, the aim is to minimize the localization error while satisfying power and illumination related constraints. However, for improved energy efficiency of VLP systems, the total power consumption of LEDs must also be taken into account in addition to localization performance requirements \cite{VLC_Efficient_2014}. Therefore, similar to the minimal illumination level problem in VLC systems \cite{Lampe_VLC_TCOM_2015,JTOD_TWC_2017}, we consider the \textit{minimum power consumption problem} for VLP systems, where the objective is to minimize the total power consumption of LEDs while keeping the CRLB of the VLC receiver below a predefined level. In the following subsections, we first investigate the problem of total power minimization under perfect knowledge of localization parameters and then study robust power allocation designs in the presence of uncertainties.\vspace{-0.3cm}

\subsection{Power Minimization with Perfect Knowledge}

In the absence of uncertainties in localization parameters, the minimum power consumption problem can be formulated as follows:\vspace{-0.3cm}
\begin{subequations}\label{eq:optPowCons}
	\begin{align} \label{eq:optPowCons_obj}
		\mathop{\mathrm{minimize}}\limits_{\pp} &~~
		\boldonet \pp
		\\ \label{eq:optPowCons_loc}
		\mathrm{subject~to}&~~
		\trace \big\{ \Jboldi(\pp) \big\} \leq \varepsilon
		\\ \label{eq:optPowCons_ind_power}
		&~~ \pp \in \mtPs
	\end{align}
\end{subequations}
where $\boldonet \pp$ determines the total electrical power consumption, $\mtPs \triangleq \mtP_1 \cap \mtP_3 \cap \mtP_4$ and $\varepsilon$ represents the maximum tolerable CRLB level for the localization of the VLC receiver. In \eqref{eq:optPowCons}, we seek to find the most energy-efficient LED power assignment scheme satisfying a certain level of localization accuracy. As implied by Lemma~1, the optimization problem in \eqref{eq:optPowCons} is convex.

The significance of the considered problem in \eqref{eq:optPowCons} for VLP systems lies in the fact that it yields the minimum value of $\Pt$ in \eqref{eq:cons_x2}, above which the optimal solution of \eqref{eq:optLED} always achieves a lower CRLB than the specified design level, $\varepsilon$. In other words, a certain level of localization performance is guaranteed by setting $\Pt$ above the obtained minimum value in \eqref{eq:optPowCons}, as in the minimal illumination level problem in VLC systems \cite{Lampe_VLC_TCOM_2015}.





\subsection{Robust Power Minimization with Imperfect Knowledge}\label{sec:worst_case}
In this part, we consider the robust counterpart of the power minimization problem in \eqref{eq:optPowCons} under deterministic norm-bounded uncertainty in matrix $\Gammabold$ in \eqref{eq:Hmatrix2} based on the error model in \eqref{eq:matrix_error_model}. Thus, we assume that the errors in $\Gammabold$ belong to a bounded uncertainty region as in Section~\ref{sec:overall_uncertainty} and develop a robust approach that guarantees the localization performance measure for all the uncertainties in the specified region. Accordingly, the robust design problem can be formulated as
\vspace{-0.1cm}\begin{subequations}\label{eq:robustPowCons_det}
	\begin{align} \label{eq:robustPowCons_det_obj}
	\mathop{\mathrm{minimize}}\limits_{\pp} &~~
	\boldonet \pp
	\\ \label{eq:robustPowCons_det_obj2}
	\mathrm{subject~to}&~~
	\trace \Big\{ \big((\Imatrix_3 \otimes \pp)^\transpose (\Gammaboldhat - \Gammabolddelta)\big)^{-1} \Big\} \leq \varepsilon \,, ~ \forall \Gammabolddelta \in \mtE
	\\ \label{eq:robustPowCons_det_gen}
	&~~ \pp \in \mtPs
	\end{align}
\end{subequations}
where $\mtE$ is given by \eqref{eq:Gamma_delta_norm} and $\varepsilon$ represents the constraint on the worst-case CRLB. Similar to \eqref{eq:optLED_robust_obj}, the semi-infinite programming problem in \eqref{eq:robustPowCons_det} can equivalently be reformulated as a convex problem, as shown in the following proposition.

\textbf{Proposition~2.} \textit{The robust power allocation problem in \eqref{eq:robustPowCons_det} is equivalent to the following convex optimization problem:}
\vspace{-0.4cm}\begin{subequations}\label{eq:optLED_robust_sdp_pow_min}
	\begin{align} \label{eq:optLED_robust_sdp_pow_min_obj}
	\mathop{\mathrm{minimize}}\limits_{\pp,\Hbold,s,\mu} &~~ \boldonet \pp
	\\ \label{eq:optLED_robust_sdp_pow_min_trace_ineq}
	\mathrm{subject~to}&~~ \trace \big\{ \Hbold \big\} \leq \varepsilon - d s
	\\ \label{eq:optLED_robust_sdp_pow_min_1}
	&~~ \Phibold(\pp,\Hbold,s,\mu) \succeq 0
	\\ \label{eq:optLED_robust_sdp_pow_min_2}
	&~~ \Hbold \succeq 0,~
	 \mu \geq 0,~
 \pp \in \mtPs
	\end{align}
\end{subequations}
\textit{where $\Phibold(\pp,\Hbold,s,\mu)$ is defined as in \eqref{eq:phi_sdp}}.

\begin{proof}
	The proof directly follows from that of Proposition~1.
\end{proof}

\section{Numerical Results}
In this section, we provide numerical examples to investigate the performance of the proposed optimal and robust power allocation designs for VLP systems.

\subsection{Simulation Setup}
We consider a VLP scenario in a room of size $10 \times 10 \times 5 ~ \rm{m}^3$, where there exist $\NL = 4$ LED transmitters and a VLC receiver. 
The locations and the orientations of the LED transmitters and the VLC receiver are provided in Table~\ref{tab:locations}. In addition, $L=4$ locations are determined for individual illumination constraints, which are also displayed in Table~\ref{tab:locations}. The average illuminance in \eqref{eq:ill_avg} is calculated over the horizontal plane of the room at a fixed height of $1\,\rm{m}$.

The scaled version of the transmitted signal, $\tildest$, in \eqref{eq:opt_var_sit} is modeled as
$\tildest = \frac{2}{3} (1-\cos({2\pi\,t}/{\Tsi})) (1+\cos(2\pi \fci \, t))$
for $i = 1,\ldots,\NL$ and $t \in \left[0, \Tsi \right]$, where $\Tsi$ is the pulse width and $\fci$ is the center frequency \cite{CRB_TOA_VLC,MFK_CRLB}.\footnote{The constant factor $2/3$ is included to satisfy $\int_{0}^{\Tsi} (\tildest)^2 dt/\Tsi = 1$, as indicated in Section~\ref{sec:opt_var}.}  From \eqref{eq:p_opt_tilde}, the optical power of $\tildest$
is calculated as $\Popttilde = 2/3$. In accordance with \cite{CRB_TOA_VLC,VLC_Guvenc_Lighting,MFK_CRLB,Lampe_MISO_TSP_2016,Light_VLC_2013}, the VLP system parameters utilized throughout the simulations are given in Table~\ref{tab:sim_params}. In addition, an asynchronous VLP system is considered, i.e., the time offsets $\{\Delta_i\}_{i=1}^{\NL}$ in \eqref{eq:tau} are modeled as unknown parameters.

\begin{table}[t]
	\centering
	\caption{Locations and Orientations}\vspace{-0.2cm}
	\begin{tabular}{ |l|l| }
		\hline
		Location of LED-$1$, $\lt{1}$  & $\left[ 1 ~ 1 ~ 5 \right]^\transpose \, \rm{m}$ \\
		Location of LED-$2$, $\lt{2}$  & $\left[ 1 ~ 9 ~ 5 \right]^\transpose \, \rm{m}$ \\
		Location of LED-$3$, $\lt{3}$  & $\left[ 9 ~ 1 ~ 5 \right]^\transpose \, \rm{m}$ \\
		Location of LED-$4$, $\lt{4}$  & $\left[ 9 ~ 9 ~ 5 \right]^\transpose \, \rm{m}$ \\
		Orientation of LEDs, $\nt{i} \, (i=1,2,3,4)$  & $\left[ 0 ~ 0 ~ -1 \right]^\transpose$ \\ 
		\hline
		Location of VLC Receiver, $\lr$  & $\left[ 3 ~ 3 ~ 0.5 \right]^\transpose \, \rm{m}$ \\
		Orientation of VLC Receiver, $\nr$  & $\left[ 0.5 ~ 0 ~ 0.866 \right]^\transpose$ \\
		\hline
		Location of Illumination Constraint-$1$, $\xx_{1}$  & $\left[ 1 ~ 1 ~ 1 \right]^\transpose \, \rm{m}$ \\
		Location of Illumination Constraint-$2$, $\xx_{2}$  & $\left[ 1 ~ 9 ~ 1 \right]^\transpose \, \rm{m}$ \\
		Location of Illumination Constraint-$3$, $\xx_{3}$  & $\left[ 9 ~ 1 ~ 1 \right]^\transpose \, \rm{m}$ \\
		Location of Illumination Constraint-$4$, $\xx_{4}$  & $\left[ 9 ~ 9 ~ 1 \right]^\transpose \, \rm{m}$ \\
		\hline
	\end{tabular}
	\label{tab:locations}
\end{table}

\begin{table}[t]
	\centering
	\caption{Simulation Parameters}\vspace{-0.2cm}
	\begin{tabular}{ |l|l| }
	\hline
	Responsivity of Photo Detector, $R_p$ & $0.4\, \rm{mA/mW}$ \\
	Area of Photo Detector, $S$ & $1\, \rm{cm} ^2$ \\
	Spectral Density Level of Noise, $\sigma^2$  & $1.3381\times10^{-22}\, \rm{W/Hz}$ \\
	LED Lambertian Order, $m_i\, (i=1,2,3,4)$  & $1$ \\
	LED Luminous Efficacy, $\kappa_i\, (i=1,2,3,4)$ & $284\,\rm{lm/W}$ \\
	Min. LED Optical Power & $5\, \rm{W}$ \\
	Max. LED Optical Power & $20\, \rm{W}$ \\
	Min. Illuminance Level, $\avgEtilde, \Etilde_{\ell} \, (\ell=1,2,3,4)$ & $30 \, \rm{lx}$ \\
	Transmitted Pulse Width, $\Tsi \, (i=1,2,3,4)$ & $1\, \mu \rm{s}$ \\
	Center Frequency, $\fci \, (i=1,2,3,4)$ & $40 + 20(i-1) \, \rm{MHz}$ \\	
	\hline		
	\end{tabular}
	\label{tab:sim_params}
\end{table}


\subsection{Power Allocation with Perfect Knowledge}

In this part, we investigate the effects of the proposed optimal power allocation approach on the localization performance of the VLC receiver under the assumption of perfect knowledge of localization related parameters. Since this is the first study to consider power allocation in VLP systems, the uniform power allocation strategy (i.e., $P_i = \Pt/\NL, \, i = 1,\ldots,\NL$) is also illustrated for comparison purposes.

\begin{figure}
	\centering
	\vspace{-0.2cm}
	\includegraphics[scale=0.6]{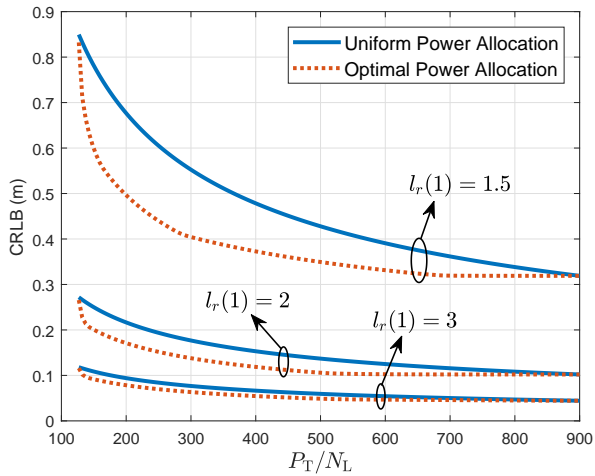}
	\vspace{-0.3cm}
	\caption{CRLB of \eqref{eq:optLED} versus $\Pt/\NL$ for optimal and uniform power allocation strategies for various locations of the VLC receiver.}\label{fig:CRLB_vs_avg_elec_power}
\vspace{-0.12in}
\end{figure}

Fig.~\ref{fig:CRLB_vs_avg_elec_power} plots the CRLB achieved by the optimal solution of \eqref{eq:optLED} versus $\Pt/\NL$, which determines the average electrical power limit, for various locations of the VLC receiver. It is observed that the optimal power allocation approach can provide significant improvements in localization performance over the conventional uniform power allocation approach. In addition, we note that the performance improvement becomes more pronounced as the VLC receiver moves away from the center of the room. The reason is that the contribution of each LED to the Fisher information in \eqref{eq:FIM_compact} becomes almost equal at the room center whereas the LEDs are less symmetric at the corners. Moreover, due to the limited linear operation regime of the LEDs, the optimal strategy exhibits a similar performance to that of the uniform strategy for sufficiently high values of $\Pt$. Furthermore, when $\Pt$ is lower than a certain value, the problem becomes infeasible due to the average illumination constraint, and the uniform and optimal strategies achieve the same CRLB at that value of $\Pt$.


\subsection{Robust Power Allocation in the Presence of Overall System Uncertainty}\label{sec:num_overall_unc}
To illustrate the performance of the robust power allocation in the presence of overall system uncertainty, as discussed in Section~\ref{sec:robust_overall}, several numerical examples are provided for the problem in \eqref{eq:optLED_robust_sdp}, which is equivalent to the original robust problem in \eqref{eq:optLED_robust_obj}. Since the goal of robustness is to optimize the worst-case performance, we investigate the worst-case CRLBs achieved by the following strategies:
\begin{itemize}
	\item \textit{Robust:} The robust strategy takes into account the uncertainty in $\Gammabold$ and solves the problem in \eqref{eq:optLED_robust_sdp}. Then, the resulting optimal value $t^{\star}$ of the slack variable $t$ yields the worst-case CRLB.
	\item \textit{Non-robust:} The non-robust strategy ignores the uncertainty in $\Gammabold$ and directly utilizes the nominal matrix $\Gammaboldhat$ in \eqref{eq:matrix_error_model} to solve the power allocation problem in \eqref{eq:optLED}. To obtain the worst-case CRLB corresponding to optimal power vector $\ppnonrobust$ of \eqref{eq:optLED}, $\ppnonrobust$ is inserted into \eqref{eq:optLED_robust_sdp} as a fixed quantity. Hence, the worst-case CRLB $t^{\star}$ is calculated by solving
	\begin{subequations}\label{eq:optLED_robust_sdp2}
		\begin{align} \label{eq:optLED_robust_sdp_obj2}
		t^{\star} = \mathop{\mathrm{min}}\limits_{t,\Hbold,s,\mu} &~~ t
		\\ \label{eq:optLED_robust_sdp_trace_ineq2}
		\mathrm{subject~to}&~~ \trace \big\{ \Hbold \big\} \leq t - d s,~ \Phibold(\ppnonrobust,\Hbold,s,\mu) \succeq 0,~
		 \Hbold \succeq 0,~
	 \mu \geq 0
		\end{align}
	\end{subequations}
	where $\Phibold(\pp,\Hbold,s,\mu)$ is given by \eqref{eq:phi_sdp}.
	\item \textit{Uniform:} In this strategy, the uniform power allocation vector is used and the corresponding worst-case CRLB is obtained via \eqref{eq:optLED_robust_sdp2} by replacing $\ppnonrobust$ with the uniform power vector.
\end{itemize}
The worst-case CRLBs are averaged over $100$ Monte Carlo realizations. For each realization, an error matrix $\Gammabolddelta$ is randomly chosen from the uncertainty set $\mtE$ in \eqref{eq:Gamma_delta_norm} and the nominal matrix $\Gammaboldhat$ is generated according to \eqref{eq:matrix_error_model}. Then, each strategy is evaluated by using realizations for which that strategy is feasible\footnote{More specifically, we fix the number of feasible realizations beforehand and continue to pick new matrices from the uncertainty region until the predefined number of feasible realizations is reached. For the robust strategy, feasibility refers to the problem in \eqref{eq:optLED_robust_sdp} being feasible for a given realization $\Gammaboldhat$. For the non-robust and uniform strategies, feasibility means that the problem in \eqref{eq:optLED_robust_sdp2} is feasible, which is equivalent to the worst-case CRLB in \eqref{eq:optLED_robust_sdp_obj2} being finite.}.

\begin{figure}
	\centering
	\vspace{-0.4cm}
	\includegraphics[scale=0.6]{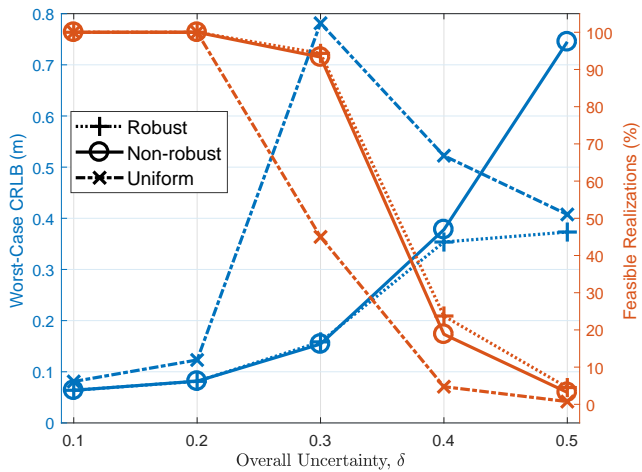}
	\vspace{-0.4cm}
	\caption{Worst-case CRLB of \eqref{eq:optLED_robust_obj} versus the level of uncertainty $\delta$, where the average power limit is $\Pt/\NL = 400$.}\label{fig:worst_case_CRLB_left_right_figure}
	\vspace{-0.4cm}
\end{figure}

Fig.~\ref{fig:worst_case_CRLB_left_right_figure} shows the worst-case CRLB performance and the feasibility rate of the considered power allocation strategies against the level of uncertainty $\delta$ in \eqref{eq:Gamma_delta_norm}. It is observed that the performance of all the strategies deteriorates as the uncertainty increases, as expected. For small uncertainty regions (i.e., small $\delta$), the robust strategy has almost the same performance as its non-robust counterpart. However, the robust strategy outperforms the non-robust strategy for large uncertainty regions, which results from the design philosophy in \eqref{eq:optLED_robust_obj}. More specifically, since the nominal matrix $\Gammaboldhat$ deviates substantially from the true matrix $\Gammabold$ for large values of $\delta$, the non-robust strategy, which treats $\Gammaboldhat$ as the true matrix in LED power optimization, results in poor worst-case localization performance. On the other hand, the robust approach attempts to minimize the performance degradation by utilizing the properties of the uncertainty region $\mtE$ in \eqref{eq:Gamma_delta_norm}.

As noted from Fig.~\ref{fig:worst_case_CRLB_left_right_figure}, the robust strategy also provides the highest feasibility rate among all the strategies since the feasible region of \eqref{eq:optLED_robust_sdp2} is smaller than that of \eqref{eq:optLED_robust_sdp} (the constraint set \eqref{eq:optLED_robust_P} is replaced by a single point in \eqref{eq:optLED_robust_sdp2}). In addition, the feasibility rate of the uniform strategy undergoes a sharp decline after a certain level of uncertainty, which distorts the monotonic behavior of its worst-case CRLB around the point where this decline occurs. It is worth noting that the non-robust strategy achieves a higher feasibility rate and lower worst-case CRLB than the uniform strategy for small $\delta$, but this trend changes as $\delta$ increases. The reason is that for small $\delta$, the non-robust approach can find near-optimal power allocation vectors in the sense of \eqref{eq:optLED_robust_obj} (since solving \eqref{eq:optLED} is almost equivalent to solving \eqref{eq:optLED_robust_obj} for small $\delta$) whereas the uniform power vector does not take into account the localization related parameters (e.g., locations and orientations of the LED transmitters and the VLC receiver) and assigns equal power to all the LEDs, which leads to low feasibility rates and large errors in localization. On the other hand, for high $\delta$, the performance of the non-robust strategy becomes worse than that of the uniform strategy with increasing errors in $\Gammaboldhat$.

%
%


\vspace{-0.2cm}
\subsection{Robust Power Allocation in the Presence of Individual Parameter Uncertainties}
In this part, we consider the robust power allocation schemes designed for the case of individual parameter uncertainties, as discussed in Section~\ref{sec:robust_ind}. In the simulations, we explore the performance of the three strategies as mentioned in Section~\ref{sec:num_overall_unc} using $100$ Monte Carlo realizations. The robust strategy is obtained by solving \eqref{eq:optLED_robust_unif_obj} via Algorithm~\ref{alg:ier}. For the non-robust strategy, the uncertainty set $\egeneralset$ in \eqref{eq:set_unif} is ignored and the nominal parameters (i.e., $\lrhat$ in \eqref{eq:lrhat} or $(\thetahat,\phihat)$ in \eqref{eq:angle_hat}) are employed for power allocation via \eqref{eq:optLED}. To compute the worst-case CRLB for a given power vector $\ppast$, which corresponds to $\Psi(\ppast)$ in \eqref{eq:func_general}, we use a multi-start optimization algorithm for globally solving the maximization problem in \eqref{eq:func_general}.

\begin{figure}
	\centering
	\vspace{-0.1cm}
	\includegraphics[scale=0.45]{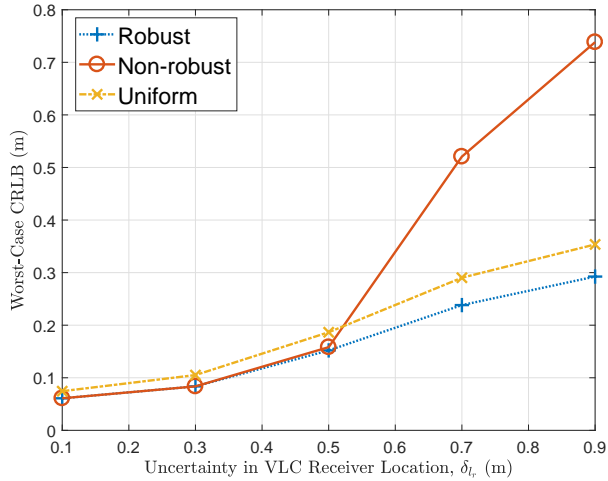}
	\vspace{-0.3cm}
	\caption{Worst-case CRLB of \eqref{eq:optLED_robust_loc_obj} versus the level of uncertainty in the location of the VLC receiver $\deltalr$, where the average power limit is $\Pt/\NL = 400$.}\label{fig:worst_case_CRLB_location}
\end{figure}

Fig.~\ref{fig:worst_case_CRLB_location} depicts the worst-case CRLB performance versus the level of uncertainty in the VLC receiver location, $\deltalr$, for the considered strategies. As seen from Fig.~\ref{fig:worst_case_CRLB_location}, the proposed robust power allocation approach always achieves lower worst-case CRLBs than the other two strategies. In addition, the performance benefit provided by the robust strategy over its non-robust counterpart becomes more evident for larger values of $\deltalr$. Hence, the robust scheme in \eqref{eq:optLED_robust_loc_obj} can effectively exploit the characteristics of the uncertainty region $\lrerset$ in \eqref{eq:error_loc} to optimize the worst-case localization performance. This also indicates that the proposed power allocation algorithm in Algorithm~\ref{alg:ier} can successfully converge to the optimal solution of \eqref{eq:optLED_robust_loc_obj}. Moreover, we observe that the uniform strategy performs worse than the non-robust strategy for small $\deltalr$. However, as $\deltalr$ increases, the non-robust approach is outperformed by the uniform approach since the latter blindly assigns equal powers to the LEDs by disregarding parameter measurements whereas the former employs the highly inaccurate measurement of $\lr$ for power allocation of the LEDs.

\begin{figure}
	\centering
	
	\subfigure[$\delta_{\phi} = 6\degree$]{\includegraphics[width=0.48\textwidth]{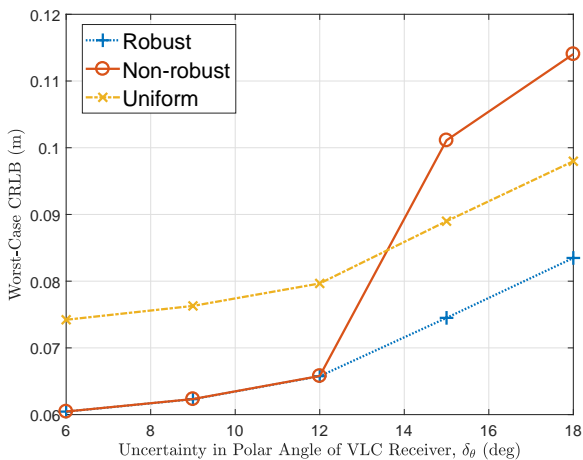}
		\label{fig:worst_case_CRLB_orientation_phi_6}
	}
	\hfill
	\subfigure[$\delta_{\phi} = 12\degree$]{\includegraphics[width=0.48\textwidth]{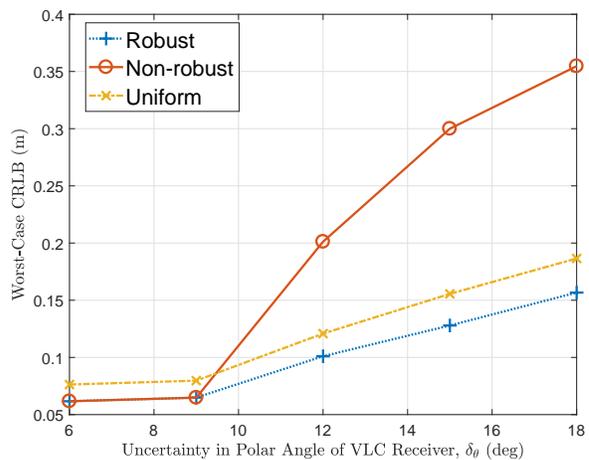}
		\label{fig:worst_case_CRLB_orientation_phi_12}
	}
	
	\caption[]{Worst-case CRLB of \eqref{eq:optLED_robust_orient_obj} versus the level of uncertainty in the polar angle of the VLC receiver $\deltatheta$ for two different values of uncertainty in the azimuth angle $\deltaphi$, where the average power limit is $\Pt/\NL = 400$. }
	\label{fig:worst_case_CRLB_orientation}
\end{figure}

In Fig.~\ref{fig:worst_case_CRLB_orientation}, we plot the worst-case CRLBs against the level of uncertainty in the polar angle of the VLC receiver ($\deltatheta$ in \eqref{eq:error_angle_theta}) for two different levels of uncertainty in the azimuth angle ($\deltaphi$ in \eqref{eq:error_angle_phi}). As seen from Fig.~\ref{fig:worst_case_CRLB_orientation}, the proposed robust power allocation strategy offers the best worst-case CRLB performance among all strategies. In addition, we note that the performance gain achieved via the robust approach becomes more prominent for larger uncertainty regions $\thetaset$ and $\phiset$ in \eqref{eq:error_angle}.



\vspace{-0.2cm}
\subsection{Minimum Power Consumption Problem}
In this subsection, numerical examples are provided for the power allocation designs in Section~\ref{sec:minPowCons}.

\subsubsection{Power Allocation with Perfect Knowledge}
We explore the electrical power consumption corresponding to the optimal solution of \eqref{eq:optPowCons} and provide a comparison with the uniform power allocation scheme, which is obtained from \eqref{eq:optPowCons_loc} as
\begin{equation}\label{eq:uniform}
P_i = \trace\big\{ \big( (\Imatrix_3 \otimes \boldone)^\transpose \Gammabold \big)^{-1} \big\} / \varepsilon
\end{equation}
for $i=1,\ldots,\NL$.

\begin{figure}
	\centering
	\vspace{-0.1cm}
	\includegraphics[scale=0.6]{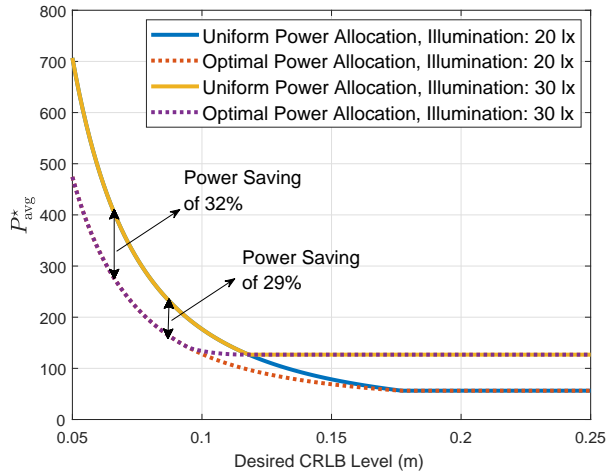}
	\vspace{-0.3cm}
	\caption{Optimal value of \eqref{eq:optPowCons_obj} divided by $\NL$ ($\Pavgstar$) versus the desired CRLB level $\sqrt{\varepsilon}$ for optimal and uniform power allocation strategies under various illumination constraints.}\label{fig:pow_vs_CRLB}
	\vspace{-0.2in}
\end{figure}

Fig.~\ref{fig:pow_vs_CRLB} plots $\Pavgstar$ versus $\sqrt{\varepsilon}$ for the optimal and uniform power allocation strategies, where $\Pavgstar$ corresponds to the optimal value of \eqref{eq:optPowCons_obj} divided by $\NL$ (which is proportional to the average electrical power consumption) and $\varepsilon$ is the desired CRLB level in \eqref{eq:optPowCons_loc}. From the figure, we observe power saving gains of around $30\%$ via the optimal approach for centimeter-level accuracy requirements. In addition, it is seen that the optimal strategy becomes equivalent to the uniform strategy when the desired level of localization accuracy is sufficiently low, which results from the illumination constraints.

\subsubsection{Robust Power Allocation with Imperfect Knowledge}
We provide examples for the case of uncertainty in VLP system parameters, discussed in Section~\ref{sec:worst_case}. For the robust strategy, we solve \eqref{eq:optLED_robust_sdp_pow_min}, which is equivalent to the original problem in \eqref{eq:robustPowCons_det} by Proposition~2, to get the optimal power vector, while the non-robust strategy is obtained by replacing $\Gammaboldhat$ with $\Gammabold$ in \eqref{eq:optPowCons}. In addition, the uniform strategy is given by \eqref{eq:uniform} with $\Gammabold$ replaced by $\Gammaboldhat$.

\begin{figure}
	\centering
	\vspace{0.1cm}
	\includegraphics[scale=0.6]{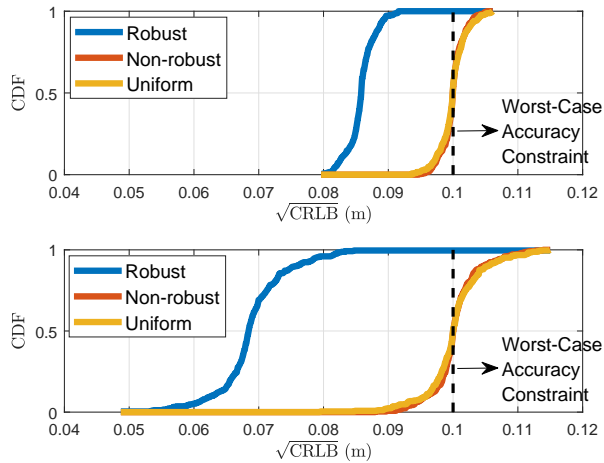}
	\vspace{-0.3cm}
	\caption{CDF of localization CRLBs achieved by robust, non-robust and uniform strategies in the case of deterministic norm-bounded uncertainty for the matrix $\Gammabold$, where the worst-case CRLB constraint in \eqref{eq:robustPowCons_det_obj2} is set to $\sqrt{\varepsilon} = 0.1\,\rm{m}$ and two different uncertainty levels are considered, namely, $\delta = 0.1$ (above) and $\delta=0.2$ (below).}\label{fig:CDF_delta_0p1_0p2}
	\vspace{-0.2cm}
\end{figure}

\begin{figure}
	\centering
	\vspace{-0.2cm}
	\includegraphics[scale=0.6]{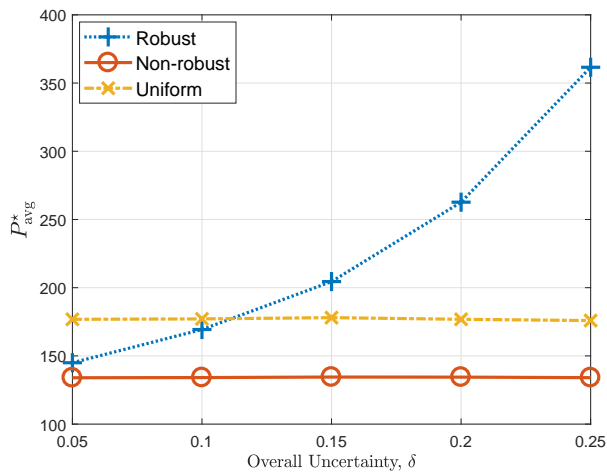}
	\vspace{-0.3cm}
	\caption{Optimal value of \eqref{eq:robustPowCons_det_obj} divided by $\NL$ ($\Pavgstar$) versus the level of uncertainty $\delta$ for robust, non-robust and uniform power allocation strategies, where the worst-case accuracy constraint is $\sqrt{\varepsilon} = 0.1\,\rm{m}$.}\label{fig:pow_cons_unc}
	\vspace{-0.12in}
\end{figure}

Fig.~\ref{fig:CDF_delta_0p1_0p2} depicts the cumulative distribution function (CDF) of the CRLBs obtained by the considered strategies for two different uncertainty levels, $\delta=0.1$ and $\delta=0.2$, by setting the worst-case accuracy level as $\sqrt{\varepsilon} = 0.1\,\rm{m}$. It is observed that the robust algorithm, which solves \eqref{eq:optLED_robust_sdp_pow_min}, satisfies the accuracy constraint in \eqref{eq:robustPowCons_det_obj2} for all the realizations of $\Gammabold$ in accordance with the robust design approach, which also verifies the validity of Proposition~2. On the other hand, the non-robust and uniform strategies are not able to satisfy the accuracy constraint for approximately $50\%$ of the realizations since they do not consider the uncertainty in $\Gammabold$ in allocating powers to the LEDs. Also, the CRLBs are observed to be more spread out for higher $\delta$ for all strategies. In Fig.~\ref{fig:pow_cons_unc}, we show $\Pavgstar$ with respect to $\delta$, where $\Pavgstar$ is the optimal value of \eqref{eq:robustPowCons_det_obj} divided by $\NL$. It is seen that the robust strategy must utilize more transmission power with increasing $\delta$ in order to guarantee the specified level of accuracy for larger uncertainty regions, as expected. Hence, the relative performance gain of the robust strategy can be achieved at the cost of higher transmit powers and increased computational complexity, which results from solving \eqref{eq:optLED_robust_sdp_pow_min} rather than the original problem \eqref{eq:optPowCons}. However, as opposed to the non-robust power allocation, the robust approach provides a solid theoretical guarantee for satisfying the worst-case CRLB constraint in \eqref{eq:robustPowCons_det_obj2}.


\section{Concluding Remarks}
In this manuscript, we have considered the problem of optimal power allocation for LED transmitters in a VLP system. The optimization problem has been formulated to minimize the CRLB for the localization of the VLC receiver under practical constraints on transmission powers and illumination levels. Under the assumption of perfect knowledge of localization related parameters, the power allocation problem has been shown to be convex and thus efficiently solvable. In the presence of overall uncertainty, we have investigated the robust design problem that aims to minimize the worst-case CRLB over deterministic norm-bounded uncertainties and proved that it can be reformulated as a convex optimization problem. In addition, we have formulated the robust min-max problems corresponding to the uncertainties in individual parameters, namely, the location and the orientation of the VLC receiver. To solve the min-max problem, we have proposed an iterative entropic regularization algorithm, whereby the original problem is transformed into a sequence of convex programs and a grid search is performed over the uncertainty region. Moreover, the problem of total power minimization has been explored under preset accuracy requirements. Simulation results have demonstrated the effectiveness of the optimal power allocation approach in enhancing the localization performance compared to the traditional uniform strategy. Furthermore, the proposed robust power allocation designs have been shown to outperform their non-robust counterparts, especially for large uncertainty regions. Regarding the minimum power consumption problem, power saving gains of $30\%$ by the optimal strategy have been observed relative to the uniform power allocation approach.

\appendix
\section{{Appendices}}

\subsection{Definition of $\gammaik$}\label{app_FIM}
$\gammaik$ in \eqref{eq:gamma_vec} is defined as follows \cite{Direct_TCOM}:
\begin{align}
\label{eq:gamma_ik}
\gammaik &=
\begin{cases}
\gammaiksyn \,, & \text{if synchronous VLP system} \\
\gammaikasyn \,, & \text{if asynchronous VLP system}
\end{cases}
\\ \label{eq:gammaiksyn}
\gammaiksyn &\triangleq \frac{R_p^2}{\sigma^2} \bigg(E_2^i\frac{\partial \alpha_i}{\partial\lrs{k_1}}
\frac{\partial \alpha_i}{\partial\lrs{k_2}}
+E_1^i\alpha_i^2\frac{\partial \tau_i}{\partial\lrs{k_1}}
\frac{\partial \tau_i}{\partial\lrs{k_2}}
-E_3^i\alpha_i\bigg(\frac{\partial \alpha_i}{\partial\lrs{k_1}}
\frac{\partial \tau_i}{\partial\lrs{k_2}}+\frac{\partial \tau_i}{\partial\lrs{k_1}}
\frac{\partial \alpha_i}{\partial\lrs{k_2}}\bigg)\bigg) \\
\gammaikasyn &\triangleq \frac{R_p^2}{\sigma^2}
\left(E_2^i-\frac{(E_3^i)^2}{E_1^i}\right)
\frac{\partial \alpha_i}{\partial\lrs{k_1}}
\frac{\partial \alpha_i}{\partial\lrs{k_2}} \\
E_1^i&\triangleq\int_{0}^{T_{s,i}}\big(\widetilde{s}'_i(t)\big)^2dt,~
E_2^i\triangleq\int_{0}^{T_{s,i}}\big(\tildest\big)^2dt,~
E_3^i\triangleq\int_{0}^{T_{s,i}}\tildest \widetilde{s}'_i(t) dt
\\\label{eq:tauDer}
\frac{\partial \tau_i}{\partial \lrs{k}}&=\frac{\lrs{k}-\lts{i}{k}}{c\norm{\lr - \lt{i}}}
\\\label{eq:alpDer}
\frac{\partial \alpha_i}{\partial \lrs{k}}&=-\frac{(m_i+1)S}{2\pi}\bigg(
\frac{\big((\lr - \lt{i})^T\nt{i}\big)^{m_i-1}}{\norm{\lr - \lt{i}}^{m_i+3}}
\big(m_i\,\nts{i}{k}(\lr - \lt{i})^T\nr
+\nrs{k}(\lr - \lt{i})^T\nt{i}\,\big)
\\\nonumber
&-\frac{(m_i+3)(\lrs{k}-\lts{i}{k})}{\norm{\lr - \lt{i}}^{m_i+5}}\big((\lr - \lt{i})^T\nt{i}\big)^{m_i}(\lr - \lt{i})^T\nr\bigg)
\end{align}
where $\widetilde{s}'_i(t)$ denotes the derivative of $\tildest$.
\vspace{-0.5cm}

\bibliographystyle{IEEEtran}
\bibliography{Ref4}

\begin{thebibliography}{10}
\providecommand{\url}[1]{#1}
\csname url@samestyle\endcsname
\providecommand{\newblock}{\relax}
\providecommand{\bibinfo}[2]{#2}
\providecommand{\BIBentrySTDinterwordspacing}{\spaceskip=0pt\relax}
\providecommand{\BIBentryALTinterwordstretchfactor}{4}
\providecommand{\BIBentryALTinterwordspacing}{\spaceskip=\fontdimen2\font plus
\BIBentryALTinterwordstretchfactor\fontdimen3\font minus
  \fontdimen4\font\relax}
\providecommand{\BIBforeignlanguage}[2]{{%
\expandafter\ifx\csname l@#1\endcsname\relax
\typeout{** WARNING: IEEEtran.bst: No hyphenation pattern has been}%
\typeout{** loaded for the language `#1'. Using the pattern for}%
\typeout{** the default language instead.}%
\else
\language=\csname l@#1\endcsname
\fi
#2}}
\providecommand{\BIBdecl}{\relax}
\BIBdecl

\bibitem{BeyondPoint}
H.~Burchardt, N.~Serafimovski, D.~Tsonev, S.~Videv, and H.~Haas, ``{VLC:
  Beyond} point-to-point communication,'' \emph{IEEE Communications Magazine},
  vol.~52, no.~7, pp. 98--105, July 2014.

\bibitem{SurveyVLC15}
P.~H. Pathak, X.~Feng, P.~Hu, and P.~Mohapatra, ``Visible light communication,
  networking, and sensing: A survey, potential and challenges,'' \emph{IEEE
  Communications Surveys \& Tutorials}, vol.~17, no.~4, pp. 2047--2077, 2015.

\bibitem{Jovicic}
A.~Jovicic, J.~Li, and T.~Richardson, ``Visible light communication:
  {O}pportunities, challenges and the path to market,'' \emph{IEEE
  Communications Magazine}, vol.~51, no.~12, pp. 26--32, Dec. 2013.

\bibitem{VLC_Survey}
D.~Karunatilaka, F.~Zafar, V.~Kalavally, and R.~Parthiban, ``{LED} based indoor
  visible light communications: State of the art,'' \emph{IEEE Communications
  Surveys \& Tutorials}, vol.~17, no.~3, pp. 1649--1678, 2015.

\bibitem{VLP_Roadmap}
J.~Armstrong, Y.~Sekercioglu, and A.~Neild, ``Visible light positioning: {A}
  roadmap for international standardization,'' \emph{IEEE Communications
  Magazine}, vol.~51, no.~12, pp. 68--73, Dec. 2013.

\bibitem{Rapid_VLC}
C.~G. Gavrincea, J.~Baranda, and P.~Henarejos, ``Rapid prototyping of
  standard-compliant visible light communications system,'' \emph{IEEE
  Communications Magazine}, vol.~52, no.~7, pp. 80--87, July 2014.

\bibitem{HighSpeedVLC}
L.~Grobe, A.~Paraskevopoulos, J.~Hilt, D.~Schulz, F.~Lassak, F.~Hartlieb,
  C.~Kottke, V.~Jungnickel, and K.~D. Langer, ``High-speed visible light
  communication systems,'' \emph{IEEE Communications Magazine}, vol.~51,
  no.~12, pp. 60--66, Dec. 2013.

\bibitem{panta2012indoor}
K.~Panta and J.~Armstrong, ``Indoor localisation using white {LEDs},''
  \emph{Electronics Letters}, vol.~48, no.~4, pp. 228--230, 2012.

\bibitem{IndoorVisLig}
H.-S. Kim, D.-R. Kim, S.-H. Yang, Y.-H. Son, and S.-K. Han, ``An indoor visible
  light communication positioning system using a {RF} carrier allocation
  technique,'' \emph{Journal of Lightwave Technology}, vol.~31, no.~1, pp.
  134--144, Jan. 2013.

\bibitem{ghassemlooy2017visible}
Z.~Ghassemlooy, L.~N. Alves, S.~Zvanovec, and M.-A. Khalighi, \emph{Visible
  Light Communications: Theory and Applications}.\hskip 1em plus 0.5em minus
  0.4em\relax CRC Press, 2017.

\bibitem{Fundamental_VLC_2004}
T.~Komine and M.~Nakagawa, ``Fundamental analysis for visible-light
  communication system using {LED} lights,'' \emph{IEEE Transactions on
  Consumer Electronics}, vol.~50, no.~1, pp. 100--107, Feb. 2004.

\bibitem{EPSILON}
L.~Li, P.~Hu, C.~Peng, G.~Shen, and F.~Zhao, ``Epsilon: {A} visible light based
  positioning system,'' in \emph{11th USENIX Symposium on Networked Systems
  Design and Implementation (NSDI)}, Seattle, WA, Apr. 2014, pp. 331--343.

\bibitem{zhang2014asynchronous}
W.~Zhang, M.~I.~S. Chowdhury, and M.~Kavehrad, ``Asynchronous indoor
  positioning system based on visible light communications,'' \emph{Optical
  Engineering}, vol.~53, no.~4, pp. 1--9, 2014.

\bibitem{LED_MultiRec}
S.-H. Yang, E.-M. Jung, and S.-K. Han, ``Indoor location estimation based on
  {LED} visible light communication using multiple optical receivers,''
  \emph{IEEE Communications Letters}, vol.~17, no.~9, pp. 1834--1837, Sep.
  2013.

\bibitem{MIMO_VLC_JSAC_2015}
K.~Ying, H.~Qian, R.~J. Baxley, and S.~Yao, ``Joint optimization of precoder
  and equalizer in {MIMO VLC} systems,'' \emph{IEEE Journal on Selected Areas
  in Communications}, vol.~33, no.~9, pp. 1949--1958, Sep. 2015.

\bibitem{MIMO_VLC_Design_TCOM_2017}
R.~Wang, Q.~Gao, J.~You, E.~Liu, P.~Wang, Z.~Xu, and Y.~Hua, ``Linear
  transceiver designs for {MIMO} indoor visible light communications under
  lighting constraints,'' \emph{IEEE Transactions on Communications}, vol.~65,
  no.~6, pp. 2494--2508, June 2017.

\bibitem{Dimming_MIMO_VLC_2016}
B.~Li, R.~Zhang, W.~Xu, C.~Zhao, and L.~Hanzo, ``Joint dimming control and
  transceiver design for {MIMO}-aided visible light communication,'' \emph{IEEE
  Communications Letters}, vol.~20, no.~11, pp. 2193--2196, Nov. 2016.

\bibitem{JTOD_TWC_2017}
Q.~Gao, C.~Gong, and Z.~Xu, ``Joint transceiver and offset design for visible
  light communications with input-dependent shot noise,'' \emph{IEEE
  Transactions on Wireless Communications}, vol.~16, no.~5, pp. 2736--2747, May
  2017.

\bibitem{Power_Offset_VLC_TCOM_2013}
K.~H. Park, Y.~C. Ko, and M.~S. Alouini, ``On the power and offset allocation
  for rate adaptation of spatial multiplexing in optical wireless {MIMO}
  channels,'' \emph{IEEE Transactions on Communications}, vol.~61, no.~4, pp.
  1535--1543, Apr. 2013.

\bibitem{ResourceAlloc_JLT_2014}
D.~Bykhovsky and S.~Arnon, ``Multiple access resource allocation in visible
  light communication systems,'' \emph{Journal of Lightwave Technology},
  vol.~32, no.~8, pp. 1594--1600, Apr. 2014.

\bibitem{PowRateOpt_VLC_TSP_2015}
C.~Gong, S.~Li, Q.~Gao, and Z.~Xu, ``Power and rate optimization for visible
  light communication system with lighting constraints,'' \emph{IEEE
  Transactions on Signal Processing}, vol.~63, no.~16, pp. 4245--4256, Aug.
  2015.

\bibitem{DCO_OFDM_TSP_2016}
X.~Ling, J.~Wang, X.~Liang, Z.~Ding, and C.~Zhao, ``Offset and power
  optimization for {DCO-OFDM} in visible light communication systems,''
  \emph{IEEE Transactions on Signal Processing}, vol.~64, no.~2, pp. 349--363,
  Jan. 2016.

\bibitem{Guvenc_JSAC_2017}
Y.~S. Eroglu, I.~Guvenc, A.~Sahin, Y.~Yapici, N.~Pala, and M.~Yuksel,
  ``Multi-element {VLC} networks: {LED} assignment, power control, and optimum
  combining,'' \emph{IEEE Journal on Selected Areas in Communications}, 2017.

\bibitem{RateOpt_VLC_JLT_2016}
R.~Jiang, Z.~Wang, Q.~Wang, and L.~Dai, ``Multi-user sum-rate optimization for
  visible light communications with lighting constraints,'' \emph{Journal of
  Lightwave Technology}, vol.~34, no.~16, pp. 3943--3952, Aug. 2016.

\bibitem{NOMA_VLC_2017}
X.~Zhang, Q.~Gao, C.~Gong, and Z.~Xu, ``User grouping and power allocation for
  {NOMA} visible light communication multi-cell networks,'' \emph{IEEE
  Communications Letters}, vol.~21, no.~4, pp. 777--780, Apr. 2017.

\bibitem{PowAlloc_VLC_JSAC_2017}
R.~Jiang, Q.~Wang, H.~Haas, and Z.~Wang, ``Joint user association and power
  allocation for cell-free visible light communication networks,'' \emph{IEEE
  Journal on Selected Areas in Communications}, 2017.

\bibitem{VLC_Precoding_TCOM_2017}
T.~V. Pham, H.~Le-Minh, and A.~T. Pham, ``Multi-user visible light
  communication broadcast channels with zero-forcing precoding,'' \emph{IEEE
  Transactions on Communications}, vol.~65, no.~6, pp. 2509--2521, June 2017.

\bibitem{PowAlloc_VLC_2012}
X.~Zhang, S.~Dimitrov, S.~Sinanovic, and H.~Haas, ``Optimal power allocation in
  spatial modulation {OFDM} for visible light communications,'' in \emph{2012
  IEEE 75th Vehicular Technology Conference (VTC Spring)}, May 2012, pp. 1--5.

\bibitem{PowAlloc_JLT_2017}
J.~Lian and M.~Brandt-Pearce, ``Multiuser {MIMO} indoor visible light
  communication system using spatial multiplexing,'' \emph{Journal of Lightwave
  Technology}, vol.~35, no.~23, pp. 5024--5033, Dec. 2017.

\bibitem{Robust_Pow_Alloc_Win_2013}
W.~W.~L. Li, Y.~Shen, Y.~J. Zhang, and M.~Z. Win, ``Robust power allocation for
  energy-efficient location-aware networks,'' \emph{IEEE/ACM Transactions on
  Networking}, vol.~21, no.~6, pp. 1918--1930, Dec. 2013.

\bibitem{Win_2014_IEEEnetw_PowerOpt}
Y.~Shen, W.~Dai, and M.~Z. Win, ``Power optimization for network
  localization,'' \emph{IEEE/ACM Transactions on Networking}, vol.~22, no.~4,
  pp. 1337--1350, Aug. 2014.

\bibitem{Win_PA_Coop_2015}
W.~Dai, Y.~Shen, and M.~Z. Win, ``Distributed power allocation for cooperative
  wireless network localization,'' \emph{IEEE Journal on Selected Areas in
  Communications}, vol.~33, no.~1, pp. 28--40, Jan. 2015.

\bibitem{Joint_Alloc_TCOM_2016}
T.~Zhang, C.~Qin, A.~F. Molisch, and Q.~Zhang, ``Joint allocation of spectral
  and power resources for non-cooperative wireless localization networks,''
  \emph{IEEE Transactions on Communications}, vol.~64, no.~9, pp. 3733--3745,
  Sep. 2016.

\bibitem{PA_Game_TSP_2016}
J.~Chen, W.~Dai, Y.~Shen, V.~K. Lau, and M.~Z. Win, ``Power management for
  cooperative localization: A game theoretical approach,'' \emph{IEEE
  Transactions on Signal Processing}, vol.~64, no.~24, pp. 6517--6532, Dec.
  2016.

\bibitem{PA_OFDM_TWC_2017}
A.~Shahmansoori, G.~Seco-Granados, and H.~Wymeersch, ``Power allocation for
  {OFDM} wireless network localization under expectation and robustness
  constraints,'' \emph{IEEE Transactions on Wireless Communications}, vol.~16,
  no.~3, pp. 2027--2038, Mar. 2017.

\bibitem{LED_Linearity}
H.~Elgala, R.~Mesleh, and H.~Haas, ``An {LED} model for intensity-modulated
  optical communication systems,'' \emph{IEEE Photonics Technology Letters},
  vol.~22, no.~11, pp. 835--837, June 2010.

\bibitem{VLC_Lighting_Mag}
J.~Gancarz, H.~Elgala, and T.~D.~C. Little, ``Impact of lighting requirements
  on {VLC} systems,'' \emph{IEEE Communications Magazine}, vol.~51, no.~12, pp.
  34--41, Dec. 2013.

\bibitem{VLC_ComMag_2014}
A.~Tsiatmas, C.~P. M.~J. Baggen, F.~M.~J. Willems, J.~P. M.~G. Linnartz, and
  J.~W.~M. Bergmans, ``An illumination perspective on visible light
  communications,'' \emph{IEEE Communications Magazine}, vol.~52, no.~7, pp.
  64--71, July 2014.

\bibitem{VLC_Guvenc_Lighting}
Y.~S. Eroglu, A.~Sahin, I.~Guvenc, N.~Pala, and M.~Yuksel, ``Multi-element
  transmitter design and performance evaluation for visible light
  communication,'' in \emph{2015 IEEE Globecom Workshops}, Dec. 2015, pp. 1--6.

\bibitem{Lighting_JLT_2008}
J.~Grubor, S.~Randel, K.~D. Langer, and J.~W. Walewski, ``Broadband information
  broadcasting using {LED}-based interior lighting,'' \emph{Journal of
  Lightwave Technology}, vol.~26, no.~24, pp. 3883--3892, Dec. 2008.

\bibitem{CRB_TOA_VLC}
T.~Wang, Y.~Sekercioglu, A.~Neild, and J.~Armstrong, ``Position accuracy of
  time-of-arrival based ranging using visible light with application in indoor
  localization systems,'' \emph{Journal of Lightwave Technology}, vol.~31,
  no.~20, pp. 3302--3308, Oct. 2013.

\bibitem{book_goldsmith}
A.~Goldsmith, \emph{Wireless Communications}, 2004.

\bibitem{multiaccessVLP}
S.~D. Lausnay, L.~D. Strycker, J.~P. Goemaere, B.~Nauwelaers, and N.~Stevens,
  ``A survey on multiple access visible light positioning,'' in \emph{IEEE
  International Conference on Emerging Technologies and Innovative Business
  Practices for the Transformation of Societies (EmergiTech)}, Aug. 2016, pp.
  38--42.

\bibitem{WirelessInfComm_97}
J.~M. Kahn and J.~R. Barry, ``Wireless infrared communications,''
  \emph{Proceedings of the IEEE}, vol.~85, no.~2, pp. 265--298, Feb. 1997.

\bibitem{Guvenc_hybrid}
A.~Sahin, Y.~S. Eroglu, I.~Guvenc, N.~Pala, and M.~Yuksel, ``Hybrid 3-{D}
  localization for visible light communication systems,'' \emph{Journal of
  Lightwave Technology}, vol.~33, no.~22, pp. 4589--4599, Nov. 2015.

\bibitem{Direct_TCOM}
M.~F. Keskin, S.~Gezici, and O.~Arikan, ``Direct and two-step positioning in
  visible light systems,'' \emph{IEEE Transactions on Communications}, vol.~66,
  no.~1, pp. 239--254, Jan. 2018.

\bibitem{Poor}
H.~V. Poor, \emph{An Introduction to Signal Detection and Estimation}.\hskip
  1em plus 0.5em minus 0.4em\relax New York: Springer-Verlag, 1994.

\bibitem{VanTrees}
H.~L.~V. Trees, \emph{Detection, Estimation, and Modulation Theory}.\hskip 1em
  plus 0.5em minus 0.4em\relax John Wiley \& Sons, New York, 2004.

\bibitem{OFDM_VLC_2014}
T.~D.~C. Little and H.~Elgala, ``Adaptation of {OFDM} under visible light
  communications and illumination constraints,'' in \emph{2014 48th Asilomar
  Conference on Signals, Systems and Computers}, Nov. 2014, pp. 1739--1744.

\bibitem{Lampe_VLC_TCOM_2015}
H.~Ma, L.~Lampe, and S.~Hranilovic, ``Coordinated broadcasting for multiuser
  indoor visible light communication systems,'' \emph{IEEE Transactions on
  Communications}, vol.~63, no.~9, pp. 3313--3324, Sep. 2015.

\bibitem{Clipping_OFDM_TCOM_2012}
L.~Chen, B.~Krongold, and J.~Evans, ``Theoretical characterization of nonlinear
  clipping effects in {IM/DD} optical {OFDM} systems,'' \emph{IEEE Transactions
  on Communications}, vol.~60, no.~8, pp. 2304--2312, Aug. 2012.

\bibitem{LED_Book}
E.~F. Schubert, \emph{Light-Emitting Diodes}.\hskip 1em plus 0.5em minus
  0.4em\relax Cambridge University Press, 2003.

\bibitem{Illuminance_Sum}
A.~Pandharipande and D.~Caicedo, ``Adaptive illumination rendering in {LED}
  lighting systems,'' \emph{IEEE Transactions on Systems, Man, and Cybernetics:
  Systems}, vol.~43, no.~5, pp. 1052--1062, Sep. 2013.

\bibitem{boyd_convex}
S.~Boyd and L.~Vandenberghe, \emph{Convex Optimization}.\hskip 1em plus 0.5em
  minus 0.4em\relax Cambridge University Press, 2004.

\bibitem{cvx2014}
M.~Grant and S.~Boyd, ``{CVX}: Matlab software for disciplined convex
  programming, version 2.1,'' \url{http://cvxr.com/cvx}, Mar. 2014.

\bibitem{Eldar_Robust_TSP_2005}
Y.~C. Eldar, A.~Ben-Tal, and A.~Nemirovski, ``Robust mean-squared error
  estimation in the presence of model uncertainties,'' \emph{IEEE Transactions
  on Signal Processing}, vol.~53, no.~1, pp. 168--181, Jan. 2005.

\bibitem{Robust_Linear_2008}
M.~B. Shenouda and T.~N. Davidson, ``On the design of linear transceivers for
  multiuser systems with channel uncertainty,'' \emph{IEEE Journal on Selected
  Areas in Communications}, vol.~26, no.~6, pp. 1015--1024, Aug. 2008.

\bibitem{Palomar_TSP_MIMO_2009}
J.~Wang and D.~P. Palomar, ``Worst-case robust {MIMO} transmission with
  imperfect channel knowledge,'' \emph{IEEE Transactions on Signal Processing},
  vol.~57, no.~8, pp. 3086--3100, Aug. 2009.

\bibitem{Robust_DRSS_TSP_2017}
Y.~Hu and G.~Leus, ``Robust differential received signal strength-based
  localization,'' \emph{IEEE Transactions on Signal Processing}, vol.~65,
  no.~12, pp. 3261--3276, June 2017.

\bibitem{LecturesConvBook}
\BIBentryALTinterwordspacing
A.~Ben-Tal and A.~Nemirovski, \emph{Lectures on Modern Convex
  Optimization}.\hskip 1em plus 0.5em minus 0.4em\relax Society for Industrial
  and Applied Mathematics, 2001. [Online]. Available:
  \url{http://epubs.siam.org/doi/abs/10.1137/1.9780898718829}
\BIBentrySTDinterwordspacing

\bibitem{boyd1994linear}
S.~Boyd, L.~El~Ghaoui, E.~Feron, and V.~Balakrishnan, \emph{Linear matrix
  inequalities in system and control theory}.\hskip 1em plus 0.5em minus
  0.4em\relax SIAM, 1994.

\bibitem{SDP_Boyd_1996}
\BIBentryALTinterwordspacing
L.~Vandenberghe and S.~Boyd, ``Semidefinite programming,'' \emph{SIAM Review},
  vol.~38, no.~1, pp. 49--95, 1996. [Online]. Available:
  \url{http://dx.doi.org/10.1137/1038003}
\BIBentrySTDinterwordspacing

\bibitem{yalmip}
J.~L{\"{o}}fberg, ``{YALMIP} : A toolbox for modeling and optimization in
  {M}atlab,'' in \emph{In Proceedings of the CACSD Conference}, Taipei, Taiwan,
  2004.

\bibitem{SDP_2010_zhi}
Z.~Q. Luo, W.~K. Ma, A.~M.~C. So, Y.~Ye, and S.~Zhang, ``Semidefinite
  relaxation of quadratic optimization problems,'' \emph{IEEE Signal Processing
  Magazine}, vol.~27, no.~3, pp. 20--34, May 2010.

\bibitem{SourceLoc_TSP_2011}
E.~Xu, Z.~Ding, and S.~Dasgupta, ``Source localization in wireless sensor
  networks from signal time-of-arrival measurements,'' \emph{IEEE Transactions
  on Signal Processing}, vol.~59, no.~6, pp. 2887--2897, June 2011.

\bibitem{Lampe_MISO_TSP_2016}
A.~Mostafa and L.~Lampe, ``Optimal and robust beamforming for secure
  transmission in {MISO} visible-light communication links,'' \emph{IEEE
  Transactions on Signal Processing}, vol.~64, no.~24, pp. 6501--6516, Dec.
  2016.

\bibitem{IterEntrReg_2004}
R.-L. Sheu and J.~Lin, ``Solving continuous min-max problems by an iterative
  entropic regularization method,'' \emph{Journal of Optimization Theory and
  Applications}, vol. 121, no.~3, pp. 597--612, 2004.

\bibitem{SmartGrid_iterativeEntReg_2014}
H.~M. Soliman and A.~Leon-Garcia, ``Game-theoretic demand-side management with
  storage devices for the future smart grid,'' \emph{IEEE Transactions on Smart
  Grid}, vol.~5, no.~3, pp. 1475--1485, May 2014.

\bibitem{VLC_Efficient_2014}
I.~Din and H.~Kim, ``Energy-efficient brightness control and data transmission
  for visible light communication,'' \emph{IEEE Photonics Technology Letters},
  vol.~26, no.~8, pp. 781--784, Apr. 2014.

\bibitem{MFK_CRLB}
M.~F. Keskin and S.~Gezici, ``Comparative theoretical analysis of distance
  estimation in visible light positioning systems,'' \emph{Journal of Lightwave
  Technology}, vol.~34, no.~3, pp. 854--865, Feb. 2016.

\bibitem{Light_VLC_2013}
J.~Gancarz, H.~Elgala, and T.~D.~C. Little, ``Impact of lighting requirements
  on {VLC} systems,'' \emph{IEEE Communications Magazine}, vol.~51, no.~12, pp.
  34--41, Dec. 2013.

\end{thebibliography}

\end{document}